\documentclass[aps,prb,superscriptaddress,nofootinbib]{revtex4}
\usepackage{graphicx}
\usepackage{array}
\usepackage{verbatim}           
\usepackage{simplewick}
\usepackage{amsmath}

\newcommand{\LL }{\left }
\newcommand{\RR }{\right }
\newcommand{\bea}{\begin{eqnarray}}
\newcommand{\eea}{\end{eqnarray}}
\newcommand{\be}{\begin{eqnarray}}
\newcommand{\ee}{\end{eqnarray}}
\newcommand{\nn }{\nonumber}

\newcommand{\D}{\mathcal{D}}
\newcommand{\dr}{\delta r}
\newcommand{\m}{{\bf m}}
\newcommand{\n}{{\bf n}}
\newcommand{\kk}{{\bf k}}
\begin{document}
\title{Dissipative Quantum Transport in Macromolecules: An Effective Field Theory Approach}
\author{E. Schneider}
\author{S. a Beccara}
\author{P. Faccioli}
\email{faccioli@science.unitn.it}

\begin{abstract}
We introduce an atomistic approach to the dissipative  quantum dynamics of  charged or neutral excitations propagating through macromolecular systems. Using the Feynman-Vernon path integral formalism, we analytically trace out from the density matrix the atomic coordinates and the heat bath degrees of freedom. This way we obtain an effective field theory which describes the real-time evolution of the quantum excitation and is fully consistent with the fluctuation-dissipation relation. The main advantage of the field-theoretic approach is that it allows to avoid using the Keldysh contour formulation. This simplification makes it straightforward to derive Feynman diagrams to analytically compute the effects of the interaction of the propagating quantum excitation with the heat bath and with the molecular atomic vibrations. For illustration purposes, we apply this formalism to investigate the loss of quantum coherence of holes propagating through a poly(3-alkylthiophene) polymer. 

\end{abstract}

\maketitle
\section{Introduction}

The investigation of the real-time dynamics of charged and neutral quantum excitations propagating through macromolecular systems is receiving a growing attention due to its potentially countless implications in nano-scale (opto-)electronics and in biophysics.   
 For example, the study of the electric conduction through inorganic~\cite{inorganic1, inorganic2, inorganic3, inorganic4, inorganic5}, organic~\cite{organic1, organic2, organic3, P3HT1}  and biological~\cite{ DNAreview, DNA_theory1, peptide} polymeric systems and aggregates is motivated by the perspective of realizing nano-scale, or even single-molecule~\cite{peptide} transistors.  
In addition, the recent experimental observation of coherent exciton transport in photosynthetic protein-pigment complexes~\cite{photo_exp1} has triggered a huge  activity aimed at clarifying the interplay between  quantum coherence, transfer efficiency and environment-driven noise~\cite{photo_theory1, photo_theory2, photo_theory3, photo_theory4}.  

Quantum transport processes in biomolecules and organic materials have  been  extensively discussed in the framework of phenomenological models 
in which the dynamics of the quantum excitation is described at the level of a one-body Hamiltonian, while the fluctuation-dissipation generated by the molecular vibrations and by the heat bath  are collectively represented by some effective bosonic fields   (see e.g. Ref.~\cite{DNA_theory1,photo_theory1,photo_theory4} and references therein). The frequency spectrum of these bosons is modeled phenomenologically, encoding  information gained from MD simulations and experiment. 
These effective models provide computationally efficient tools to study the global and general mechanisms which underlie the long-range charge transport in macromolecules and investigate de-coherence and re-coherence phenomena. 

In order to  establish a more direct connection between the quantum transport dynamics and the specific physico-chemical properties of the molecule under consideration, an alternative approach~\cite{DNA_theory1,DNA_theory3, P3HT1} has been developed in which the atomic coordinates are treated explicitly and are evolved in time using a classical MD algorithm. The parameters of the effective tight-binding Hamiltonian are derived from the electronic structure, hence depend on the instantaneous molecular configuration. They are evaluated at periodic time intervals,  within the Born-Oppenheimer approximation, e.g  in the  DFT-TB scheme, while the current flowing through the system  is  estimated in Landauer theory.
This method neglects the effect of the coupling between the quantum excitation and the atomic nuclei on the molecular dynamics.  In addition, the charge is assumed to propagate instantaneously across the system. In general, the validity of these assumptions may be questionable whenever the quantum transfer dynamics and the molecular dynamics occur at comparable  time scales.  

A theoretical framework which takes into account of all the correlations between the quantum excitation dynamics and the molecular dynamics  is the so-called \emph{non-equilibrium Green's function} (NEGF) method (see e.g. Ref.~\cite{NEGF1} and references therein). This formalism was  introduced to describe electric conduction across  molecular wires and has been extensively applied to systems consisting of a few hundreds of atoms. In the NEGF approach, the interaction with the metallic leads and with phonons is treated at the quantum level, since the method is often used to investigate conduction  at very low temperatures. The real-time dynamics of the system is described in the Keldysh formalism by a Dyson equation containing different types of Green's functions, corresponding to different segments of the Keldysh contour. Unfortunately, this feature makes the approach quite cumbersome and computationally demanding, hence hardly applicable to large molecular systems,  such as DNA or 
of  
conjugate polymers.

Fortunately, when investigating the transport of electric charge or neutral quantum excitations across macromolecules at room temperature, a quantum description of the molecular vibrations is not really mandatory, as it is demonstrated by the success of classical molecular dynamics (MD) simulations.  On the other hand, in these studies it is important to take into account of the fluctuation and dissipation generated by the solvent (in biomolecules) or by neighboring molecules ( in organic frameworks ). The NEGF approach does not exploit the possibility of taking the classical limit for the molecular vibrations and does not explicitly account for the fluctuation-dissipation effects induced by the heat bath. 
 
In the present paper, we  develop a formalism to describe quantum transport in macromolecular systems at room temperature, in which we exploit the possibility of treating the molecular vibrations at the classical level. In analogy with the NEGF method,  the equations of motions for the entire system are not postulated phenomenologically, but rather rigorously derived from the reduced density matrix of the system. This way, one retains in a consistent way the relevant couplings between the molecular, heat bath and quantum charge degrees of freedom. 

The computational efficiency of our approach is enhanced by the fact that we are able to analytically integrate out the vibronic and heat bath dynamics and obtain a simpler  (but rigorous) effective theory, which is formulated solely in terms of the quantum excitation degrees of freedom. 
In this theory, 
fluctuation-dissipation effects and thermal oscillations of the molecule are taken into account through effective interaction terms, which are derived from first principles. The information about the configuration-dependent electronic structure of the molecule is implicitly encoded in the 
parameters which appear in the effective theory. These are 
determined once and for all by means of quantum-chemistry calculations. 

A second important feature of  our theory is that it allows to strongly simplify the formalism to describe the non-equilibrium dissipative dynamics. Indeed, our final path integral representation of the  density matrix is \emph{formally} equivalent to that of a vacuum-to-vacuum Green's function in a zero-temperature  quantum-field theory. In particular, the time variable in the action functionals is integrated along the real axis, and not along the Keldysh contour, like in the NEGF approach. 
The formal connection with quantum field theory in vacuum makes it straightforward to derive Feynman diagrams to perturbatively compute the effects of the dissipative coupling between the propagating quantum excitation,  the heat bath and atomic degrees of freedom. These  corrections are obtained analytically, i.e. without any need to average over many independent MD trajectories. 

As a first illustrative application of this formalism, we develop a simple model to describe intra-chain charge propagation in a  poly(3-alkyothiophene) (P3HT). First, all the couplings in the effective theory are derived  from quantum chemistry calculations, at the  DFT-TB level. Next,   the leading-order perturbative corrections to the evolution of the charge density distribution are obtained by computing a few Feynman diagrams. The results are then compared with those obtained by solving numerically the coupled quantum/stochastic equations of motion for the polymer configurations and the charge reduced density matrix. From this comparison, we are able to assess the range of parameters and time intervals over which we expect the perturbative approach to be reliable. We also monitor the progressive loss of quantum coherence and identify the effective interactions which are responsible for de-coherence and re-coherence phenomena. 

The manuscript is organized as follows:
The formalism and path integral representation of the density matrix is developed in sections \ref{Hamiltonian} and \ref{FV}. In section \ref{PTheory} we derive the perturbation theory and compute the relevant-leading order Feynman diagrams. In section \ref{model} we present our illustrative application to the P3HT system.  Conclusions and perspective developments are summarized in section \ref{conclusions}. 
 
\section{Modeling the Dynamics of Quantum Excitations in Macromolecules }
\label{Hamiltonian}

An entirely {\it ab-initio} approach to quantum transport processes in macromolecules would involve solving the complete time-dependent Sch\"odinger equation which couples all the nuclear and electronic degrees of freedom, both in the molecule and in its surrounding environment. Clearly, such an approach is beyond the reach of any present or foreseen computational scheme.  

A commonly adopted framework to reduce the computational complexity of this problem consists in
relying on the Born-Oppenheimer approximation and in coarse-graining of the electronic problem into an effective configuration-dependent one-body tight-binding Hamiltonian. 
In such an approach, the molecule is first partitioned into several fragments, hereby labeled by an index $\m$, each of which is assigned a frontier orbital $|\phi_{\bf m} \rangle$. Such a partition of the molecule  must be defined in such a way that the electron density is significantly more delocalized within each molecular fragment than over different fragments~\cite{Quantum_Chemistry0}. The frontier orbitals $|\phi_{\bf m} \rangle$ are calculated by solving the Schr\"odinger equation for a reduced portion of the molecule, centered at  the fragment ${\bf m}$. 
The system's wave function is then obtained by diagonalizing the Hamiltonian projected onto the space of the frontier orbitals  $|\phi_{\bf m} \rangle$.

For example, in studying electronic hole transport in DNA, the molecular fragments can be chosen to coincide with the base-pairs and their highest occupied molecular orbitals (HOMO's) can be obtained by solving the Schr\"odinger equation for an isolated pair~\cite{DNA_theory2}.  The propagation of the charge carriers through the molecule is then modeled as the hopping of holes between neighboring base-pairs. 

\begin{figure}[t!]
\begin{center}
\includegraphics[width=7cm]{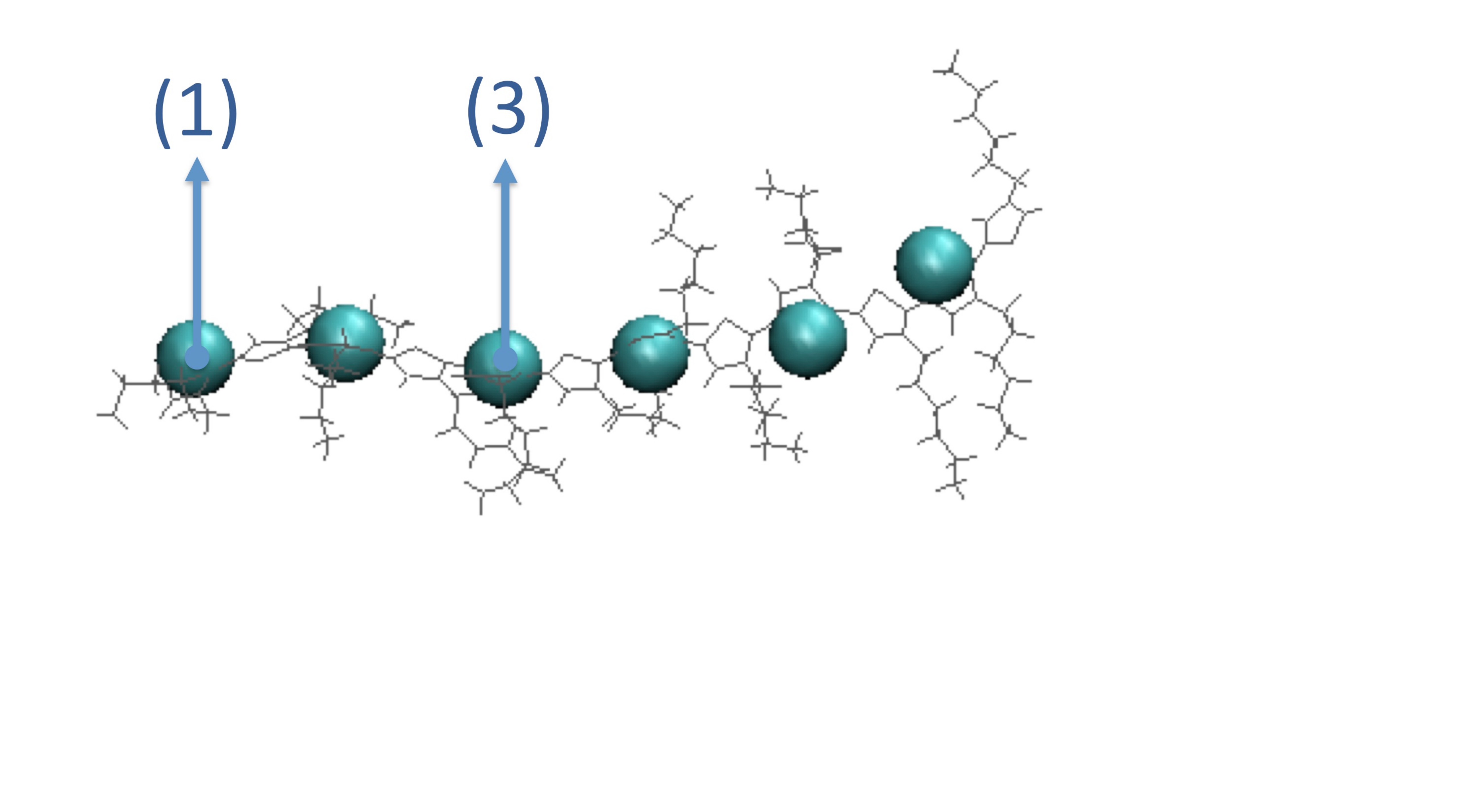}
\includegraphics[width=5cm]{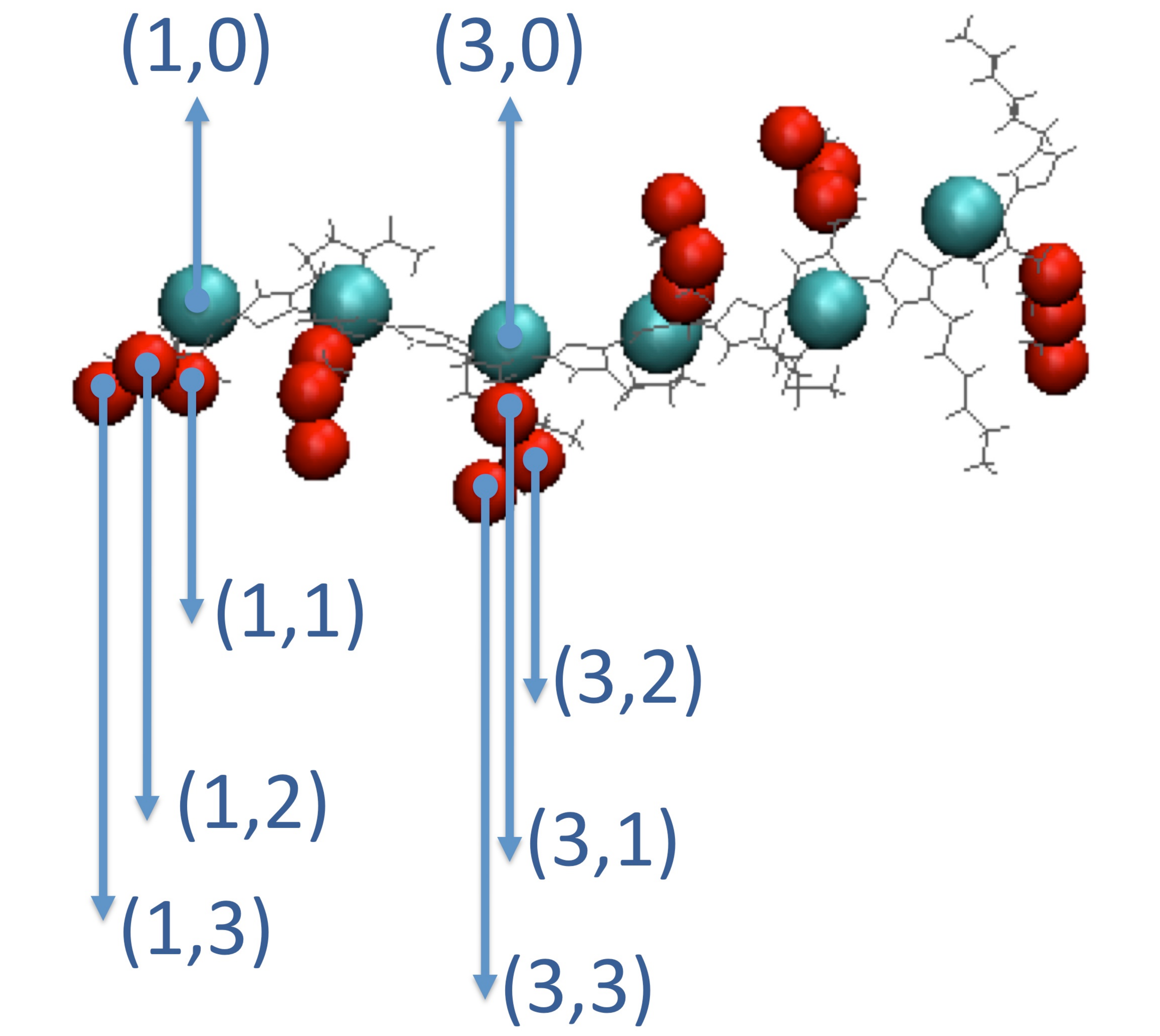}
\includegraphics[width=5cm]{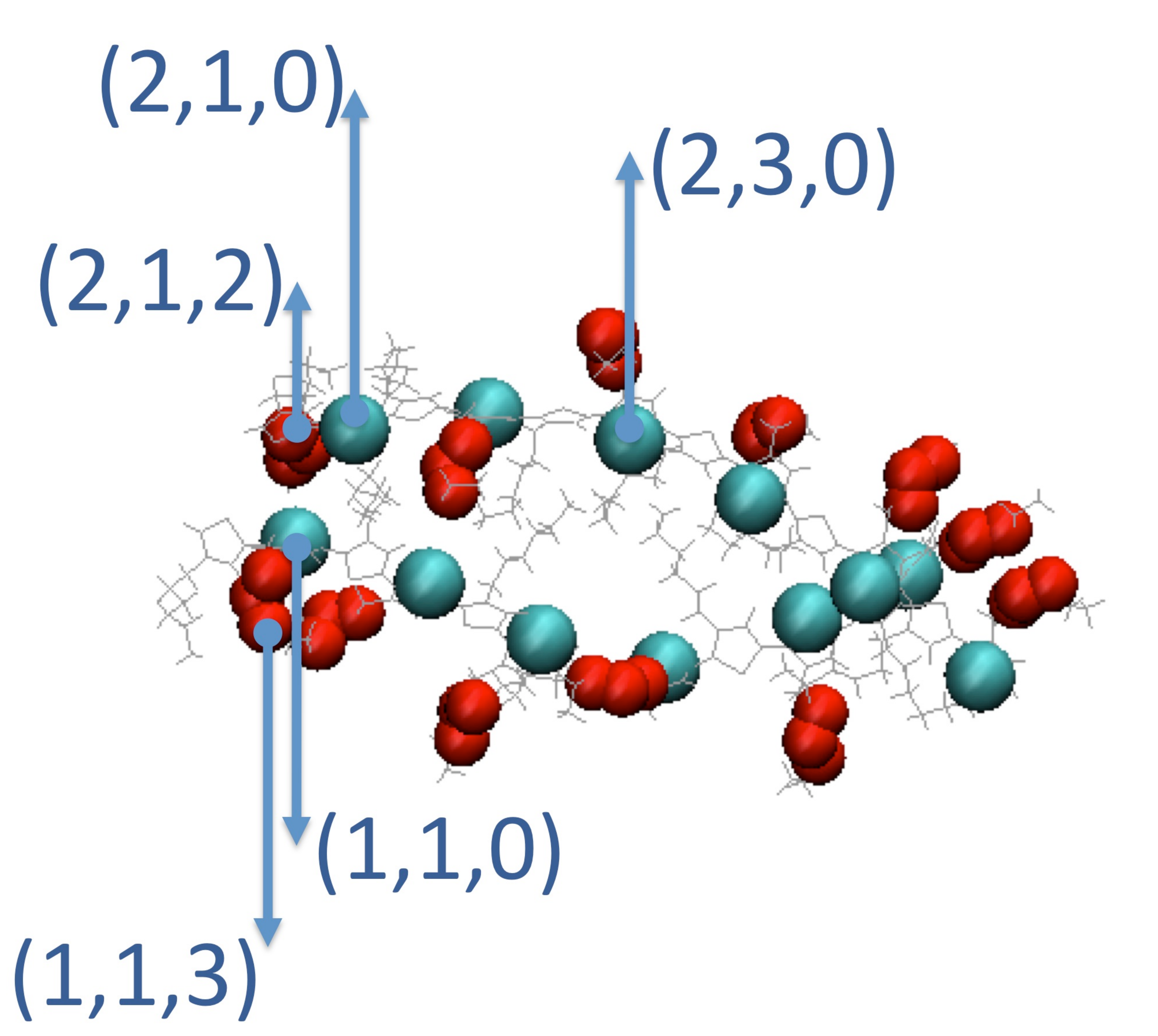}
\caption{Labeling of molecular fragments in molecules with  different topology. In linear polymers, the fragments among which the charge hops form an effective 
one-dimensional system
 ( left panel). In branched polymers, a two dimensional representation may be introduced in order to account for propagation along the side chains (center panel). In polymer assemblies a third index may be used 
to label the different molecules (right panel). }
\label{dimFig}
\end{center}
\end{figure}

In general, it is convenient to choose the dimensionality of the molecular fragment index $\m$ according to the topology of the  system under consideration. For example, in molecular wires it is natural to adopt a one-dimensional index (see left panel in Fig.~\ref{dimFig}).
On the other hand, if the electron  can also propagate along the side-chains of branched polymers, a two-dimensional index is most appropriate (see center panel in Fig.~\ref{dimFig}). In assemblies of branched polymers, a third component of the index $\bf m$ may be introduced  to distinguish between the different monomeric units.
 
In the following, we shall denote with $r^i=(r_x^{i},r_y^{i},r_z^{i})$ the Cartesian coordinates of the $i-$th atom. The set of all atomic nuclear coordinates is collectively represented by the configuration space vector $Q$,
\be
Q (t)\equiv (q_1,\ldots, q_{3N}) = (r^{1}_{x}(t),r^{1}_{y}(t),r^{1}_{z}(t);\ldots;
r^{N}_{x}(t),r^{N}_{y}(t),r^{N}_{z}(t)). 
\ee  

The dynamics of the entire system is modeled by the following quantum Hamiltonian:
\be
\label{Htot}
\hat H = \hat H_{MC} + \hat H_M + \hat H_B + \hat H_{MB}.
\ee
In this equation,   $\hat H_{MC} $ the tight-binding Hamiltonian for the quantum excitation,
\be
\label{HMC}
\hat H_{MC} = \sum_{s=1,2}\sum_{\m,\n=1} f_{\m \n}[Q]~\hat a^\dagger_{\m, s} \hat a_{\n,s }.
\ee  
We stress that the transfer matrix elements depend on the molecular configuration $Q$. The $\hat a^\dag_{\m, s}$ ($\hat a_{\m, s}$) operators create (annihilate) a quantum excitation at the molecular fragment $\bf m$ and $s$ denotes the third component of the spin. For sake of definiteness, in this work we shall focus on the propagation of electron holes in Highest Occupied Molecular Orbitals (HOMO's), which is relevant for many  molecular charge transfer phenomena.  Correspondingly, the creation and annihilation operators are assumed to obey the anti-commutation relations
\be
\{\hat a_{\m, s}, \hat a_{\n,s'}\}=\{\hat a^\dag_{\m,s}, \hat a^\dag_{\n, s'}\}=0,\qquad \{\hat a_{\m, s}, \hat a^\dag_{\n, s'}\} = \delta_{\bf l m}~\delta_{s s'}.
\ee
The generalization of the present formalism to the case in which the propagating excitation is bosonic (e.g. an exciton) is straightforward and will not be discussed explicitly.  
Depending on the specific molecular system,  it may be necessary to include  also the coupling between different molecular orbitals. This can be done by introducing additional creation-annihilation operators. 

Sometimes it is convenient to split the $f_{\m \n}[Q]$ matrix elements into the hopping matrix elements $T_{\m \n}[Q]$ and on-site energies $e_{\m}[Q]~\delta_{\m \n}$:
\be \label{HOPPING_matrix}
f_{\m \n}(Q) \equiv T_{\m \n}(Q)(1-\delta_{\m \n}) - e_{\bf n}(Q)\delta_{\m \n}.
\ee
The parameters  $T_{\m \n}$ and $e_{\m}$ are obtained from the fragment orbitals  $|\phi_{\m }\rangle$ and  $|\phi_{\n}\rangle$ and depend parametrically on the configuration vector $Q$:  
\be\label{TFO1}
 T_{\m \n}(Q) &\equiv& \langle \phi_{\m} |\hat{\mathcal{H}}_{el.}|\phi_{\n} \rangle,\\
 \label{TFO2}
 e_{\m}(Q) &\equiv& \langle \phi_{\m} | \hat{\mathcal{H}}_{el.}|\phi_{\m} \rangle,
 \ee 
where $\hat{\mathcal{H}}_{el}$ is the electronic Hamiltonian (for example, the Kohn-Sham Hamiltonian of density functional theory).  
In a molecular wire, the Hamiltonian $\hat H_{MC}$  must include also the coupling with the leads which play the role of electron source and sink.

The Hamiltonian $\hat H_M$ in Eq. (\ref{Htot}) governs the conformational dynamics 
of the atomic nuclei\footnote{Notice that, for sake of notational simplicity, we are assuming that all atoms have the same mass $M$. The generalization to different atomic masses is straightforward and will not be discussed here. }, in the absence of electronic holes and reads
\be
\label{HM}
\hat H_M \equiv \frac{\hat{P}^{2}}{2 M} + \hat V(Q),
\ee
where the $P$ is the momentum canonically conjugated to the configuration vector $Q$.  $V(Q)$ is the molecular 
 potential energy, evaluated in the Born-Oppenheimer approximation. This includes the interaction between the different atoms within the molecule and possibly with the external fields.  We stress that the potential energy $V(Q)$ in Eq. (\ref{HM}) depends only on the molecular configuration. This is equivalent to taking the adiabatic limit for the dynamics, and to assume that the location of the quantum excitation does not alter  in a significant way the interaction between the atomic nuclei. In many cases of interest, the validity of this approximation has been questioned and corrections to the adiabatic regime have been proposed. However, in this first paper we shall not deal with these complications. 

The part of the Hamiltonian $\hat H_B + \hat H_{MB}$ describes
the coupling of the molecule with a thermal heat bath in the Leggett-Caldeira model~\cite{legget-caldeira}, i.e. through an infinite set of harmonic-oscillators coupled to each atomic coordinate: 
\be
\hat H_{B} &=& \sum_{\alpha=1}^{3 N}\sum_{j=1}^{\infty}\left(\frac{\hat \pi_j^2}{2\mu_j}+\frac{1}{2}\mu_j \omega_j^2 \hat x_j^2  \right) \, ,\\
\label{HMB}
\hat H_{MB} &=& \sum_{\alpha=1}^{3 N}\sum_{j=1}^{\infty}\left(- c_j \hat x_j \hat q_\alpha + 
~ \frac{c_j^2}{2\mu_j \omega_j^2}\hat q_\alpha^2\right)\, .
\ee
$X=(x_1, x_2, \ldots)$ and $\Pi = (\pi_1, \pi_2, \ldots)$ are the harmonic oscillator coordinates and momenta, 
 $\mu_j$ and $\omega_j$ denote their masses and 
 frequencies and $c_j$  are the couplings between atomic and heat bath variables.   The last term in Eq. (\ref{HMB})  is a  standard counter-term introduced to compensate the renormalization of the molecular potential energy which occurs when the heat bath variables
  are traced out (see e.g. the discussion in Ref.~\cite{PIreview}).

The  model introduced so far represents the starting point of many approaches which have been proposed to describe quantum transport in molecular systems.  For example, in the method used in Ref.~\cite{P3HT1} and Ref.~\cite{DNA_theory2}, the evolution of the molecular degrees of freedom $Q$ is described at the classical level by means of a MD algorithm based on the force fields obtained by parametrizing the molecular potential energy $V(Q)$. Then, the current  flowing  through the molecule is evaluated at regular time intervals,  in the Landauer formalism, using the instantaneous values of the $f_{\bf l m}[Q(t)]$ tight-binding coefficients.
Such an approach retains the effects of the molecular thermal oscillations on the charge dynamics. Indeed, fluctuations on $Q$ generate dynamical disorder in the tight-binding matrix elements $f_{\bf l m}(Q)$. On the other hand, this approach neglects the back-reaction of the charge dynamics on the molecular dynamics. For example, it does not account for the fact that the system's total energy may be decreased by visiting  some specific atomic configurations in which some on-site energies of the quantum excitation are lower. 

In a recent paper~\cite{boninsegna}, the path integral formalism was used to explicitly eliminate the dynamics of the heat bath variables, and  take the classical limit for the atomic degrees of freedom. As a result, an algorithm was derived to describe in a dynamical way the coupled evolution of the molecule and the charge in the heat bath. In particular, in the high-friction limit, the quantum and stochastic evolution of the charge wave function $|\Psi\rangle $ and the molecular configuration $Q$
in an infinitesimal  time interval $\Delta t$ is determined by the following equations:
\be \label{EoMs}
\left\{
\begin{array}{rl}
& q_\alpha(t+\Delta t) = q_\alpha(t) - \frac{\Delta t}{M \gamma} \frac{\partial}{\partial q^\alpha}\left(V[Q(t)]
+ \sum_{\bf m n}  [ \rho_{\bf m n}(t)~f_{\bf n m} (Q )]  \right) + \sqrt{2 \frac{k_B T\Delta t}{M \gamma}} \xi_\alpha(t) \\
&\\
&|\Psi(t+\Delta t) \rangle = e^{-\frac{i \Delta t }{\hbar}~ \hat H_{MC}[Q(t)]}  | \Psi(t)\rangle,
\end{array}
\right.
\ee
where $\rho_{\bf l m}(t) = \langle \Psi(t) | {\bf l} \rangle \langle {\bf m} | \Psi(t) \rangle$ is the time-dependent (reduced) density matrix of the 
hole, $\gamma$ is the heat bath friction coefficient and  $\xi_a(t)$ is a stochastic variable sampled from a Gaussian distribution with zero average and unitary 
variance.  Since the motion of the molecular degrees of freedom is stochastic, the charge probability density at different times is obtained from the average over many 
independent trajectories, which may turn out to be a computationally challenging procedure. 

In the next section, we review and further develop the path integral approach to hole-transport in macromolecules and obtain a scheme which does not require to perform any MD or Langevin simulation. 


\section{Effective Field Theory for the  Reduced Density Matrix} \label{FV}

Let us assume that a hole is  initially created at the HOMO of some molecular fragment $\kk_{i}$. We are interested in computing the conditional probability  $P_t(\kk_f,|\kk_i)$ for the hole to be found at the HOMO of the molecular fragment $\kk_f$, after a time interval $t$.  Such a probability is described by the following time-dependent reduced density matrix element: 
\be
\label{Pcond}
P_t(\kk_f| \kk_i) &=& \frac{\text{Tr} [~|\kk_f \rangle \langle \kk_f| \hat \rho(t)~]}{\text{Tr} ~\hat \rho(t)} =
\frac{\text{Tr} [ |\kk_f \rangle \langle  \kk_f| 
 e^{-\frac{i}{\hbar}\hat H t}~\hat \rho(0)~e^{\frac{i}{\hbar}\hat H t}]}
 {\text{Tr}~ \hat \rho(0)},\qquad
 \ee
where  $\hat \rho(0)$ is the initial density matrix, which is taken to be in the factorized form
\be 
\label{factor}
\hat \rho(0) =  |  \kk_i \rangle \langle \kk_i| ~\times ~e^{-\frac{1}{K_B T}\hat H_{M}}~\times ~e^{-\frac{1}{K_B T}\hat H_{B}}.
\ee 
Eq. (\ref{factor}) corresponds to assuming that, at the initial time, the molecular degrees of freedom and the heat bath degrees of freedom can be considered 
separately equilibrated at the same temperature. 
Hence, the normalization factor at the denominator reads
\be
\text{Tr}~ \hat \rho(0) = \text{2} \cdot Z_{M}(\beta)~ \times Z_{B}(\beta),
\ee 
where $Z_{M}$ and $Z_B$ are the  quantum partition functions for the molecule and heat bath degrees of freedom and the degeneracy factor 2 follows from enumerating the initial spin states. 
We emphasize that the ratio in Eq.~(\ref{Pcond}) expresses dynamics of all the degrees of freedom at the fully quantum level.

\begin{figure}[t!]
\begin{center}
\includegraphics[width=10cm]{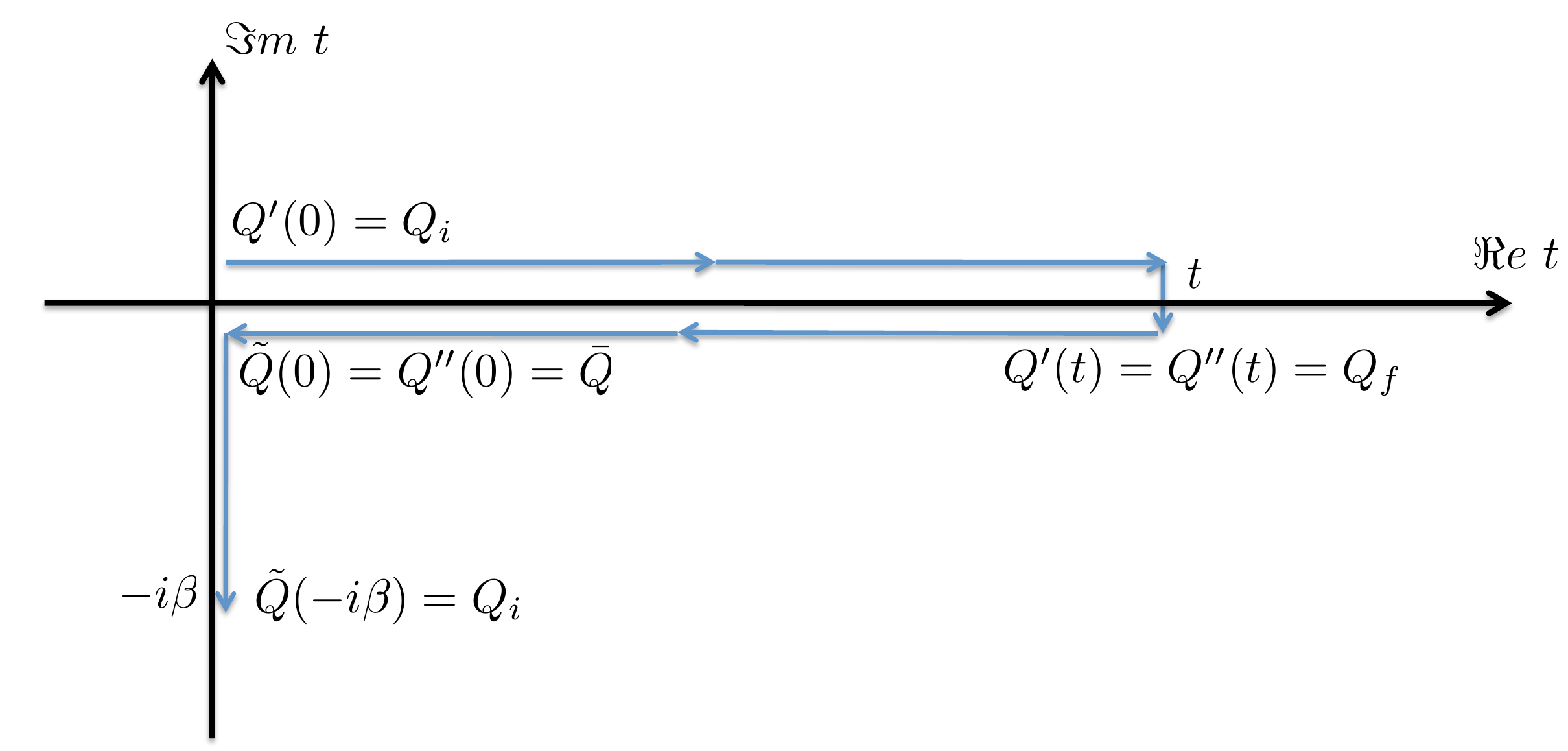}
\caption{Boundary conditions of the molecular configurations paths defined on the Keldysh contour appearing in the path integral (\ref{PI2}). }
\label{Keld}
\end{center}
\end{figure}

Let us now derive a path integral expression for its numerator. 
To this end, we choose to adopt a second-quantized representation\footnote{As we shall see below, the field-theoretic description for the quantum charge is adopted in order to obtain a simpler final representation of the density matrix. Indeed, it allows to replace the time integration along the Keldysh contour (shown in Fig.~\ref{Keld}) with a standard time integral along the real axis.} of the quantum charge dynamics, while retaining the standard first-quantized representation for the dynamics of the
  atomic coordinates and for the harmonic oscillators in the heat bath. 
  
The path integral representation of the reduced density matrix (\ref{Pcond}) is obtained by performing the Trotter decomposition of the  forward- and backward- time evolution operators $e^{-i \hat H t}$ and $e^{i \hat H t}$ and of the imaginary-time evolution  operators $e^{-\frac{1}{K_B T}\hat H_{M}}$ and $e^{-\frac{1}{K_B T}\hat H_{B}}$ which appear in Eq.s~(\ref{Pcond}) and (\ref{factor}). 
In practice, our choice of the representation of the charge, the heat bath and the molecule dynamics corresponds to introducing the following resolution of the identity:
\be
1 &=&  \int d Q \int d X \int \left( \prod_{\stackrel{\kk}{ s=1,2}}  \frac{d\phi_{\kk, s} d\phi_{\kk,s}^*}{2 \pi i}\right)
e^{-\sum_{l=1}\phi_{{\bf l}, s} \phi^*_{{\bf l}, s} } |Q, X, \Phi \rangle~\langle Q, X, \Phi |,
 \ee
where the states $|Q, X, \Phi \rangle$
 collect the set of all hole's coherent states (constructed from the annihilation operators associated to each molecular fragment,  $\hat a_{\kk, s}$), the set of the eigenstates of the molecular coordinate operator $\hat Q$ and the set of eigenstates of the heat bath generalized coordinate operator $\hat X$.  
 Throughout this paper we shall adopt Einstein's notation and implicitly assume the summation over all repeated indexes, except for the initial and final position of the molecule, ${\bf k}_i$ and ${\bf k}_f$, which are held fixed. 

  Once the conditional probability (\ref{Pcond})  is written in the path integral form and the Gaussian functional integral over the harmonic oscillator variables is carried out analytically, one reaches the expression
\be
\label{PI2}
P_t(\kk_f~| \kk_i) &=&\frac{1}{2 Z_M(\beta)}
\int dQ_f \int dQ_i \int d\bar Q \int_{\bar Q}^{Q_i} \D  \tilde Q ~e^{-\beta S_{E}[\tilde Q]}~ \int_{Q_i}^{Q_f}\D Q^{'}  \int^{\bar Q}_{Q_f} \D Q'' \int \D \phi^{'} \D \phi^{'*}  \int\D \phi'' \D \phi^{''*} \nn\\
&\times & e^{- \phi^{' *}_{\m}(0)\phi^{'}_{\m}(0)}~e^{-\phi^{''*}_{\m}(t)\phi^{''}_{\m}(t)}  
~(\phi^{'}_{\kk_f}(t)\phi^{' *}_{\kk_i}(0)~ \phi^{'' *}_{\kk_f}(t)\phi^{''}_{\kk_i}(0))~e^{-\Phi[Q',Q'']} \nonumber\\
&\times & e^{\frac{i}{\hbar}S_{MC}[Q^{'},\phi^{'},\phi^{'*}]-\frac{i}{\hbar} S_{MC}[Q^{''},\phi^{''},\phi^{''*}]}\, ,
\ee 
where $\int dQ_f$, $\int dQ_i$ denote  the standard (i.e. Riemann) integral  over the final and initial molecular configurations, respectively,  $\int d \bar Q$ 
is an integral over some intermediate configuration and $\mathcal{D}Q$ denotes the functional integral measure (see Fig.~\ref{Keld}). The functionals which appear at the exponent are defined as 
\be
S_{E}[\tilde Q] &=& \int_0^\beta d\tau ~\frac{M}{2} \dot{\tilde{Q}}^2(\tau)+ \int_0^\beta d\tau  V[\tilde Q(\tau)] \, , \\
S_{MC}\left[Q, \phi, \phi^{*}\right]  &=&  \int_0^t dt'~\frac{M}{2} \dot Q^2(t')-\int_0^t dt'~\left\{V[Q(t')]
+   \phi_{\m}^*(t')\left( i\hbar\frac{\partial }{\partial t'}\delta_{\m \n} -f_{\m \n}[Q(t')] \right) \phi_{\n}(t')\right\} \, , \\
\Phi[Q',Q'']&=&\frac{1}{\hbar}\int_0^t dt' \int_0^{t'} dt'' \left\{ \left(Q'(t')-Q''(t') \right)\cdot\left[\mathcal{B}(t'-t'')Q'(t'') - \mathcal{B}^*(t'-t'')Q''(t'') \right]\right\}\nonumber\\
&+&  i\frac{\bar\mu}{2\hbar}\int_0^t dt'\left[{Q'}^2(t')-{Q''}^2(t') \right], \qquad  \left(\bar\mu = \frac{c_j^2}{m_j \omega_j^2}\right) \, . 
\ee
$\mathcal{B}(t)$ is a Green's function which encodes the fluctuation-dissipation induced by the heat bath and reads:
\be
\mathcal{B}(t) =  \frac{c_j^2}{\mu_j \omega_j}\left[\text{coth}\left(\frac{\omega_k \hbar}{2 k_B T}\right)~\text{cos}(\omega_j t)- i ~\text{sin}(\omega_j t)\right].
\ee
The  time scales at which thermal oscillations are damped and memory effects in the heat bath are relevant can be tuned by varying the
 frequencies of the virtual harmonic oscillators in Eq.  (\ref{HMB}). In particular, here we consider the so-called Ohmic bath limit,
 in which the  $\mathcal{B}(t)$ reduces to 
 \be
 \label{Lohm}
\mathcal{B}(t) \rightarrow \mathcal{B}^{ohm}(t) =\frac{2 k_B T M \gamma}{\hbar} \delta(t)+ \frac{ i~ M \gamma}{2}~ \frac{d}{dt}\delta(t).
 \ee
It is well known that, in the classical limit for the molecular motion, this choice for $\mathcal{B}(t)$ leads to a Langevin dynamics with friction coefficient $\gamma$  (see e.g. Ref.~\cite{PIreview}).  Hence, Eq. (\ref{PI2}) represents a quantum generalization of the stochastic dynamics of the molecule, which includes also the time-evolution of the hole.  As usual, the path integral (\ref{PI2}) is defined over Grassmann fields or complex fields, depending if the propagating excitation is a fermion or a boson. 

 It is important to note that the $Q^{'}(t)$ and $Q^{''}(t)$ variables which appear in the path integral (\ref{PI2}) represent the configuration of the molecule propagating forward and backwards in time respectively, while the $\tilde Q(t)$ variables are associated to the evolution of the same molecule along the imaginary time direction. All these paths can be collectively represented by path variable integrated along the so-called Keldysh contour (see Fig.~\ref{Keld}). 
  
Equivalently, the path integral (\ref{PI2}) can be expressed in a  form in which  forward and backward molecular paths $Q'(t)$ and $Q''(t)$ are replaced by their average and difference, respectively:
 \be
 R=\frac{1}{2}\left(Q^{'} + Q^{''}\right)\, , \qquad y=Q^{'}-Q^{''} \, .
 \ee
The result is
\bea \label{PIy}
P_t(\kk_f | \kk_i)&\equiv& \frac{1}{2 Z_M(\beta)} \int \mathcal{D}\phi^{'} \mathcal{D}\phi^{'\ast} \phi^{'}_{\kk_f}(t) \phi^{'\ast}_{\kk_i}(0) ~e^{- \phi^{'*}_{\m}(0)\phi^{'}_{\m}(0)} 
~e^{\frac{i}{\hbar} S_{0}[\phi^{'}, \phi^{'*}]}\nonumber\\
&\times&\int \mathcal{D}\phi^{''} \mathcal{D}\phi^{''\ast}~ \phi^{''\ast}_{\kk_f}(t) \phi^{''}_{\kk_i}(0)~ e^{-\phi^{''\ast}_{\m}(t)\phi^{''}_{\m}(t)}  e^{-\frac{i}{\hbar}  S_{0}[\phi^{''}, \phi^{''*}]}\nonumber\\
&\times&\int dQ_f \int d\bar Q\int dQ_i \int_{\bar Q}^{Q_i} \D \tilde Q~e^{-S_{E}[\tilde Q]}~  \int_{\frac{1}{2}(Q_i+\bar Q)}^{Q_f} \mathcal{D}R \int_{Q_i-\bar Q}^{0} \mathcal{D}y~e^{\frac{i}{\hbar}\left(\mathcal{W}[R,y,  \phi^{'\ast},\phi^{'}, \phi^{''*},\phi^{''}]
+ i \hbar ~\Phi^{'}[R,y]\right)} \, , \nn \\
\eea
where the functionals at the exponent are defined as
\bea
\mathcal{W}[R, y, \phi^{'*},\phi^{'}, \phi^{''*},\phi^{''}] &=& \int_0^t dt' \left\{M~ \dot R \cdot \dot y - V\left[R + \frac{y}{2}\right]
 +  V\left[R - \frac{y}{2}\right] \right.\nonumber\\
&-&\left. \left[f_{\n \m}\left[R+\frac{y}{2}\right]~\phi^{' *}_{\n} \phi^{'}_{\m} -f_{\n \m}\left[R-\frac{y}{2}\right]~\phi^{'' *}_{\n} \phi^{''}_{\m} \right] \right\} \, , \\
S_{0}[\phi,\phi^{*}]&=&  \int_0^t dt'  \phi_{\n}^\ast(t') \left[ i\hbar~ \frac{\partial}{\partial t'}\delta_{\n \m}
 \right]\phi_{\m}(t') \, .
\eea

In the path integral representation (\ref{PI2}), the time-evolution of the charge-molecule system in contact with a dissipative heat bath follows directly from the quantum Hamiltonian defined in  Eq. (\ref{Htot}), without any further approximation. In particular,   the molecule's configurational dynamics and the charge's quantum propagation are described at the fully quantum level.  In addition, there is no restriction either on the strength of the coupling between the molecular vibrations  and the charge, nor on  the amplitude of the  conformational changes which the molecule can undergo within the time interval $t$. Clearly,  computing such a path integral is a formidable task and further approximations are needed. 

Our \emph{first approximation} consists in taking the classical limit for the dynamics of the molecular atomic coordinates. To this end, we begin by noting that the saddle-point equations which are derived by functionally differentiating  the exponent in Eq.~(\ref{PI2}) with respect to $R, y, \phi', \phi''$ lead to the condition $y(t)=0$ for all $t$ (see derivation in Ref.~\cite{boninsegna}). Following the discussion in Ref.~\cite{PIreview}, we impose the classical limit on the molecular motion  by assuming that  the path $y(t)$ remains close to its saddle-point configuration (hence represents a small fluctuation) and by imposing the boundary-condition $y(0)=0$.

We can verify that the correspondence principle is fulfilled by such an approximation. Indeed, in the absence of the quantum charge, the resulting expression for the conditional probability coincides with the Onsager-Machlup path integral representation of the Langevin dynamics of the molecule in its heat bath. To prove this, let us provisorily drop all the coherent fields, expand the functionals in the exponents to quadratic order in $y$  and perform the resulting Gaussian integration. This way, the path integral reduces to    
\be
\frac{1}{Z_M(\beta)} ~\int dQ_f \int dQ_i \int_{Q_i}^{Q_i} \D \tilde Q  e^{-S_{E}[\tilde Q]}~\int_{Q_i}^{Q_f} \mathcal{D} R ~e^{-S_{OM}[R]}=1, 
\ee
where $S_{OM}[R]$ is the well-known Onsager-Machlup functional, which  assigns a statistical weight to the stochastic trajectories in the Langevin dynamics:
\be
S_{OM}[R]=\frac{\beta}{ 4 M \gamma } \int_0^t d t' ~\left[ M \delta \ddot{R} + \frac{\partial}{\partial R} V(R ) 
 + M \gamma \delta \dot{R}\right]^2 .
\ee
Finally, we take the saddle-point approximation for the path integral $\D \tilde Q$, which corresponds to taking the classical limit also for the partition function of  the initial molecular configuration. The result is
\be
\int dQ_f \int dQ_i \frac{e^{-\beta V[Q_i]}}{Z_M[\beta]}~\int_{Q_i}^{Q_f} \mathcal{D} R ~e^{-S_{OM}[R]} = \int dQ_f \int dQ_i
~P_t(Q_f, Q_i)~\rho_0(0) = 1 \, ,
\ee
where $\rho_0(Q_i) = \frac{e^{-\beta V[Q_i]}}{Z_M[\beta]}$ is the initial distribution of molecular configurations.  We recognize that this is the normalization condition on the solution of the Fokker-Planck equation, expressed in path integral form~\cite{DRP3, DRP4} and is the starting point of the so-called dominant-reaction pathway approach to investigate the long-time dynamics of macromolecules~\cite{DRP1, DRPFIP}. 

Let us now return to the path integral expression in Eq.~(\ref{PIy}), in  which a quantum charge is allowed to propagate across the molecule and discuss our \emph{second approximation}. We note that the quantum transport dynamics is in general much faster than the characteristic time scales for major conformational transitions of macromolecular systems (typically ranging from few nanoseconds to many seconds or even larger). Hence, during the time intervals which are relevant for quantum propagation phenomena, the molecule can be assumed to follow at most only small oscillations around the mechanical equilibrium configuration $Q_0$, which is defined as the global minimum of the molecular potential energy $V(Q)$. In this small-oscillation limit, it is convenient to introduce the atomic displacement variables
 \be
\delta r(t) = R(t)-Q_0 \, ,
 \ee
 and regard both the $\delta r(t)$  and  $y(t)$ as small quantities of the same order.
 
In the expansion of the  $\mathcal{W}$ functional up to quadratic order in $\delta r$ and $y$  we obtain a term 
\bea 
V\left(R-\frac{y}{2}\right) - V\left(R+\frac{y}{2}\right) & = & \frac{1}{2} \mathcal{H}_{ij} \left[ \left(\dr - \frac{y}{2}\right)_i \left(\dr - \frac{y}{2}\right)_j\
 - \left(\dr + \frac{y}{2}\right)_i \left(\dr + \frac{y}{2}\right)_j \right] \nn\\
 &=& \dr_i y_j \mathcal{H}_{ij} + \ldots \, ,
\eea
where $\mathcal{H}_{i j}\equiv \frac{\partial}{\partial Q_i\partial Q_j} V(Q)\Large |_{Q=Q_0}$ is the Hessian matrix of the potential energy at the point of mechanical 
equilibrium.

A small deviation from the equilibrium configuration $Q_0$ generates a small change in the hopping matrix elements  and in the on-site energies which define the tight-binding Hamiltonian (\ref{HMC}). To the leading-order in the Taylor expansion  in powers of $\dr$ and $y$ we have:
\be
\label{E1}
f_{\n \m}\left(r-\frac{y}{2}\right) &=& f_{\n \m}^0 + f_{\n \m}^i \left( \dr_i - \frac{y_i}{2} \right) + \ldots \, ,\\ 
\label{E2}
f_{\n \m}\left(r+\frac{y}{2}\right) &=& f_{\n \m}^0 + f_{\n \m}^i \left( \dr_i + \frac{y_i}{2} \right) + \ldots \, , 
\ee
where  
$
f^0_{\n \m} \equiv f_{\n\m}(Q_0)$ and 
$
f_{\n \m}^i \equiv \frac{\partial}{\partial Q^i} f_{\n \m}(Q)|_{Q=Q_0}.
$
In the small oscillation regime, the path integral over $y$  is of Gaussian type and can be performed analytically, yielding 
\be
\label{PIstep3}
&&P_t(\kk_f,t | \kk_i) =\frac{1}{2 Z_M(\beta)}\int \mathcal{D}\phi^{'} \mathcal{D}\phi^{'\ast} ~\phi^{'}_{\kk_f}(t) \phi^{'\ast}_{\kk_i}(0)~ e^{-\phi^{'*}_{\m}(0)\phi^{'}_{\m}(0)} 
e^{\frac{i}{\hbar} S_{0}[\phi^{'}, \phi^{'\ast}]}\nonumber\\
&&\times~\int \mathcal{D}\phi^{''} \mathcal{D}\phi^{''\ast} ~\phi^{''\ast}_{\kk_f}(t) \phi^{''}_{\kk_i}(0) ~e^{-\phi^{''*}_{\m}(t)\phi^{''}_{\m}(t)}  e^{-\frac{i}{\hbar} S_{0}[\phi^{''}, \phi^{''*}]}~ \nn\\
&& \times ~\int \mathcal{D} \delta r ~e^{\frac{i}{\hbar}S[\delta r, \phi^{'\ast},\phi^{'},\phi^{''\ast},\phi^{''}]} e^{-\frac{\beta}{2} \delta r_i(0)\mathcal{H}_{i j} \delta r_j (0)}, \nn\\
\ee
where the functional integral over $ \delta r(t)$ is unrestricted also at time $0$ and time $t$ and the $S$ functional reads
\bea \label{1step}
 \mathcal{S}[\dr, \phi^{'\ast},\phi^{'},\phi^{''\ast},\phi^{''}] & = & \frac{ i \hbar ~\beta}{ 4 M \gamma } \int_0^t d t' ~\left[ M \delta \ddot{r}_i + \mathcal{H}_{ij} \dr_j 
 + M \gamma \delta \dot{r}_i + \frac{1}{2} f_{\n \m}^i (\phi_{\n}^{'\ast} \phi_{\m}^{'} +\phi_{\n}^{''\ast}\phi_{\m}^{''} ) \right]^2  \nn \\ 
  &-&    \int_0^t d t' ~\left( f^0_{\n \m} + f^i_{\n \m} \dr_i \right) (\phi_{\n}^{'\ast} \phi_{\m}^{'} -\phi_{\n}^{''\ast}\phi_{\m}^{''} ). 
 \eea

It is convenient to introduce a differential operator $ \hat L$, which depends on  molecular coordinate indexes $i, j$: 
\be
[\hat L]_{ij} =   M \left( \partial_{t'}^2 + \gamma \partial_{t'} \right)\delta_{ij} + \mathcal{H}_{ij} \, ,
\ee
its Hermitian conjugate reads
\be
[\hat L^\dag]_{ij} =  M \left( \partial_{t'}^2 - \gamma  \partial_{t'}\right) \delta_{ij} + \mathcal{H}_{ij},
\ee
Note that in $\hat L$ and $ L^\dag$ the time-derivatives are defined to act on the right.
Using such operators, the functional in Eq.~(\ref{1step}) can be rewritten as 
\bea
&&\hspace{-1cm}\mathcal{S}[\dr,\phi', \phi^{'\ast}, \phi'',\phi^{''\ast}]   =  \frac{i \hbar ~\beta }{4 M \gamma } \int_0^t d t'  ~\dr_i(t') [ \hat L^{\dag}\cdot \hat L]_{ij} ~\dr_j (t') \nn \\
 &+& \frac{i \hbar~ \beta }{4 M \gamma }~ \int_0^t d t' ~ f_{\n \m}^i~ (\phi_{\n}^{'\ast}(t') \phi_{\m}^{'}(t') +\phi_{\n}^{''\ast}(t')\phi_{\m}^{''}(t') )
   ~ [\hat L]_{ij} ~\dr_j (t') \nn \\
   &+&  \frac{i \hbar \beta }{16 \gamma M }\int_0^t d t'  ~ f_{\n\m}^i  ~ (\phi_{\n}^{'\ast}(t') \phi_{\m}^{'}(t')+\phi_{\n}^{''\ast}(t')\phi_{\m}^{''}(t') )~
  (\phi_{\bf l}^{'\ast}(t') \phi_{\bf h}^{'}(t') +\phi_{\bf l}^{''\ast}(t')\phi_{\bf h}^{''}(t') )~  f_{\bf l h}^i.  \nn\\
 &-&  \int_0^t dt'~   f^0_{\n \m}
 ~(\phi_{\n}^{'\ast}(t') \phi_{\m}^{'}(t') -\phi_{\n}^{''\ast}(t')\phi_{\m}^{''}(t') )\nn\\
 &-& \int_0^t dt'~f_{\n \m}^ i ~(\phi_{\n}^{'\ast}(t') \phi_{\m}^{'}(t') -\phi_{\n}^{''\ast}(t')\phi_{\m}^{''}(t') )~  \dr_i (t').
 \eea
 
The path integral (\ref{PIstep3}) describes the coupled dynamics of the nuclear coordinates and of the electronic hole in the molecule. 
It is instructive to first consider this conditional probability in the limit in which the couplings between 
the charge and the molecular degrees of freedom are completely neglected. In this case, the path integral factorizes as 
\be\label{decoupled}
P_t(\kk_f | \kk_i) &\simeq& |G^>_0(\kk_f,t| \kk_i)|^2 ~\times~\left\{
\frac{1}{ Z_M(\beta)} \int \mathcal{D}\delta r~e^{-\frac{\beta}{4 M \gamma} \int_0^t d\tau \left(M \delta \ddot{r} - \gamma\delta{\dot r} + \mathcal{H}_{i j} \delta r_j \right)^2} e^{-\frac{\beta}{2} \delta r_i(0)\mathcal{H}_{i j} \delta r_j (0)}\right\}\\
&=&|G^>_0(\kk_f,t| \kk_i)|^2.
\ee
In this equation,  
\be 
G^>_0(\kk_f, t| \kk_i) &\equiv& \langle \kk_f | e^{-\frac{i}{\hbar} \hat H_0 t}~|~\kk_i, \rangle, \qquad \text{with}\quad
\hat H_{0} \equiv ~f_{\bf l m}(Q_0) ~\hat{a}^\dag_{\bf l}~\hat{a}_{\bf m} \, ,
\ee
is the hole's Green's function defined by the tight-binding Hamiltonian $\hat H_0$ , which is evaluated keeping the molecule  ``frozen" in its minimum-energy configuration $Q_0$.

\subsection{Dirac-Like Notation}

Let us now return to the most general case, in which the interactions between the hole, the molecule and the heat bath are fully taken into account. The symmetric structure of the exponent in the path integral representation of the conditional probability $P(\kk_f,t|\kk_i)$  
suggests to collect all coherent field degrees of freedom 
$\phi^{'}_{\uparrow {\bf n}},\phi^{'}_{\downarrow {\bf n}}, \phi^{''}_{\uparrow {\bf n}}, \phi^{''}_{\downarrow {\bf n}}$ into a single 4-compenent Grassmann field $\psi$ defined as:
\be\label{notation1}
\psi_{\bf n} &\equiv& \left(
\begin{array}{c}
\phi^{'}_{{\bf n}, \uparrow}\\
\phi^{'}_{{\bf n}, \downarrow}\\
\phi^{''}_{{\bf n},\uparrow }\\
\phi^{''}_{{\bf n},\downarrow}\\
\end{array}
\right).\ee
Similarly, we collect all conjugate fields into
$
\psi^\dagger_{\bf n} \equiv  \left(\phi^{'\ast}_{{\bf n}, \uparrow},\phi^{'\ast}_{{\bf n}, \downarrow}, \phi^{\ast ''}_{{\bf n}, \uparrow},\phi^{\ast ''}_{{\bf n}, \downarrow} \right).
$
In view of the formal analogy with Dirac  theory\footnote{Clearly,  in the case in which the propagating particle is a scalar boson,  the $\psi$ field has only scalar field two-components, $\psi=(\phi', \phi'')$  and the gamma matrixes are replaced by Pauli matrices $\gamma_0 \to \tau_3$, $\gamma_5\to \tau_1$.} it is convenient to introduce also the following $4\times 4$ matrixes, which define the projection onto the upper and lower spinor components and the interchange between them:
\be
\label{notation2}
\gamma_0 = \left(
\begin{array}{cccc}
1 & 0 & 0 & 0 \\
0 & 1 & 0 & 0 \\
0 & 0 & -1 & 0 \\
0 & 0 & 0 & -1
\end{array}
\right), 
\qquad
\gamma_+ = \left(
\begin{array}{cccc}
1 & 0 & 0 & 0 \\
0 & 1 & 0 & 0 \\
0 & 0 & 0 & 0 \\
0 & 0 & 0 & 0
\end{array}
\right), 
\qquad
\gamma_- = \left(
\begin{array}{cccc}
0 & 0 & 0 & 0 \\
0 & 0 & 0 & 0 \\
0 & 0 & 1 & 0 \\
0 & 0 & 0 & 1
\end{array}
\right),
\qquad 
\gamma_5 = \left(
\begin{array}{cccc}
0 & 0 & 1 & 0 \\
0 & 0 & 0 & 1 \\
1 & 0 & 0 & 0 \\
0 & 1 & 0 & 0
\end{array}
\right).
\ee
In addition, we change variable in the integration over the $\psi^\dagger$ field by means of the substitution  
\be
\label{notation3}
\bar \psi_{\bf n}(t) \equiv \psi_{\bf n}^\dag(t) ~\gamma_0.
\ee

Using the notation defined in Eq.s (\ref{notation1}), (\ref{notation2}) and (\ref{notation3}), the conditional probability is  written as:
\be
\label{PIstep4}
P_t(\kk_f | \kk_i) &\equiv& \frac{(-1)}{2 Z_M(\beta)}
 ~\int \D \delta r~\mathcal{D}\bar{\psi}~ \D \psi~ e^{-\mathcal{L}_1(t,0)-\mathcal{L}_2(t,0)}
 \left(\bar \psi_{\kk_f}(t) ~\gamma_- \gamma_5~  \psi_{\kk_f}(t)~ \bar{\psi}_{\kk_i}(0)~\gamma^+ \gamma_5~\psi_{\kk_i}(0)\right)  \nn\\
&\times&~\exp\left(\frac{i}{\hbar} S_0 [\bar \psi, \psi]\right)~\exp\left(-S_{eff.}[ \delta r]\right)~ \exp\left(\frac{i}{\hbar}\left\{I_1[\delta r, \bar \psi, \psi]+I_2[\delta r, \bar \psi, \psi] + I_3[\bar \psi, \psi]\right\}\right) \, ,
\ee
where the terms
\be
S_{0}[\bar \psi, \psi] &=& \int_0^t dt'~\bar{\psi}_{\bf m} ~( i \hbar \partial_{t'} - f^0_{\bf m n})~\psi_{\bf n} \, , \\
S_{eff}[\delta r] &=& \frac{\beta }{4 M \gamma } \int_0^t d t'  ~\delta r^i~ [ \hat L^{\dag}\cdot \hat L]_{i j}~\delta r^j \, , 
\ee
describe the  quantum propagation of the charge and the Langevin dynamics the molecular coordinates, in the absence of any coupling between them. The functionals $I_1$, $I_2$ and $I_3$ are the interaction terms and are defined as
\be
&&I_1[\delta r, \bar \psi, \psi]= -\int_0^t dt' ~J^i~\delta r^i
\\
&&I_2[\delta r, \bar \psi, \psi]=\frac{i \hbar \beta }{4 M \gamma }~ \int_0^t d t'~  J^i_0~ [\hat L]_{i j} ~\delta r^j, \\
&&I_3[\bar \psi, \psi] = \frac{ i \hbar \beta }{16 \gamma M }\int_0^t d t'  J^i_0~J^i_0 \, ,
 \ee
with 
$
J^i   \equiv ~\bar{\psi}_{\bf m} ~f^i_{\bf m n}~\psi_{\bf n}$ and $
J_0^i \equiv ~\bar{\psi}_{\bf m} ~\gamma_0~f^i_{\bf m n}~\psi_{\bf n}.
$
We note that the couplings $I_1$ and $I_{2}$ vanish in the classical limit  $\frac{\hbar \beta}{M \gamma} \to 0$.  The surface terms $\mathcal{L}_1(t,0)$ and $\mathcal{L}_2(t,0)$ follow from  the over completeness of the coherent-field basis and from the Boltzman distribution of the initial configuration, respectively, and read
\be
\mathcal{L}_1(t,0)&=&\left(\bar{\psi}_{\m}(0)~\gamma_0~\gamma^+ \psi_{\m}(0) + \bar{\psi}_{\m}(t)~~\gamma_0~\gamma_-~\psi_{\m}(t)\right) \, ,\\
\mathcal{L}_2(t,0)&=& \frac{\beta}{2} \delta r_i(0)\mathcal{H}_{i j} \delta r_j (0) + \ldots \, 
\ee

Some comments on Eq. (\ref{PIstep4}) are in order. Firstly, we note that the overall minus sign appearing in front the integral is a consequence of the Fermi statistics and ensures the overall positivity of the probability density. Secondly, we observe that, while the path integral  (\ref{PIstep3}) is defined over forward- and backward- propagating fields (i.e. along the Keldish  contour), the path integral Eq. (\ref{PIstep4}) contains only the integration in the forward time direction. Indeed, the backward-propagating fields have been replaced by  lower-components of the ``spinor" field, hence can be \emph{formally} interpreted as anti-matter degrees of freedom propagating forward in time. 

A major simplification which follows from our approximations is that the integral over the small displacement of the molecular coordinates from their equilibrium  position $\delta r$ can be performed analytically. The result is an effective theory with non-instantaneous interactions between the holes:
\be\label{PIstep5}
P_t(\kk_f | \kk_i) &\equiv& -\int \mathcal{D}\bar{\psi}~ \D \psi~ e^{-\mathcal{L}_1(t,0)} ~ \left(\bar \psi_{\kk_f}(t) ~\gamma_- \gamma_5~  \psi_{\kk_f}(t)~ \bar{\psi}_{\kk_i}(0)~\gamma^+ \gamma_5~\psi_{\kk_i}(0)\right)~e^{-\frac{i}{\hbar}S_0[\bar\psi, \psi]}\nn\\
&&\times~e^{\frac{i}{4\hbar} \int_0^t d t'~dt''~J^i_0(t') \mathcal{V}_{i j}(t'-t'') J^i(t'')- \frac { M \gamma }{ \beta \hbar^2 } \int_0^t dt' dt''J^i(t') \Delta_{i j}(t'-t'') J^i(t'')} \, .
\ee
In this equation,  $\Delta_{ij}(t'-t'')$, $\mathcal{V}_{ij}(t-t')$ are respectively the Green's functions of the $[\hat L^\dag \hat L]$ operator and the sum of the Green's functions of the $\hat L$ and $\hat L^{\dag}$ operators.

In order to explicitly compute them, it is convenient to transform to the normal mode basis, in which the Hessian of the potential energy at the minimum energy configuration $Q_0$ is diagonal:
  \be \label{Matrix_diagonal}
  U^\dag_{k s}~\mathcal{H}_{s j} U_{j i} = \delta_{k i} ~M~\Omega^2_k.   
  \ee
In this equation,  $\Omega_k$ denotes the frequency of the $k$-th normal mode. 

In this basis, the expressions of the vibron propagators $\Delta_{ij}(t'-t'')$ and  $\mathcal{V}_{ij}(t-t')$ read (see the derivation in the appendix \ref{AppA}) 
\be
\label{Delta}
\Delta_{i j}(t) &=& 
 \frac{e^{-\frac{1}{2}\gamma~|t|}}{2~M^2~\Omega_k^2}~ U^\dag_{i k}~U_{k j}~a_{k}(t) \, ,\\
\label{DeltaV}
\mathcal{V}_{i j}(t) &=& \frac{2~e^{-\frac{1}{2}\gamma~|t|}}{M~\omega_0^{k}}~ U^\dag_{i k}~U_{k j}~b_{k}(t) \, ,
\ee
where $\omega_0^k=\sqrt{\LL|4 \Omega_k^2 - \gamma^2 \RR|}$ and
\bea
a_k(t)&=&\left\{ 
\begin{array}{cc}
 \displaystyle{ \frac{\sin  \LL( \frac{1}{2} \omega_0^k~|t| \RR)}{\omega_0^k}  +  \frac{\cos  \LL( \frac{1}{2} \omega_0^k~|t| \RR)}{\gamma} }  & \quad  \text{if} ~ 2 \Omega_k \geq \gamma \,,  \\ 
 \displaystyle{ \frac{\sinh \LL( \frac{1}{2} \omega_0^k~|t| \RR)}{\omega_0^k}  +  \frac{\cosh \LL( \frac{1}{2} \omega_0^k~|t| \RR)}{\gamma} }  & \quad \text{if} ~ 2 \Omega_k < \gamma \, , 
\end{array}
\right.\\
b_k(t) & = &
\left\{
\begin{array}{cc} 
 \quad \sin \LL( \frac{1}{2} \omega_0^k~|t| \RR)  & \quad \qquad  \text{if} ~ 2 \Omega_k \geq \gamma \, ,\\
 \quad \sinh \LL( \frac{1}{2} \omega_0^k~|t| \RR) & \quad \qquad  \text{if} ~ 2 \Omega_k < \gamma \, .\\
\end{array} 
\right.\\
\eea

It is useful to consider the asymptotic expressions for the Green's functions  in the limit $\gamma \gg 2 \Omega_k$, which corresponds to the over-damping regime:
\be
\label{Delta_over}
\Delta_{i j}(t) 	&=& \displaystyle{ \frac{e^{-\frac{\Omega_k^2}{\gamma}~|t|}}{2 ~M^2~\gamma~\Omega_k^2} ~U^\dag_{i k}~U_{k j}} \, ,\\
\label{DeltaV_over}
\mathcal{V}_{i j}(t) 	&=& \displaystyle{ \frac{1}{M~\gamma} \LL(1 + 2 \frac{\Omega_k^2}{\gamma} \RR)\, \LL( e^{ - \frac{2 \Omega_k^2}{\gamma} |t|} - e^{- \gamma |t|} \, e^{ \frac{2 \Omega_k^2}{\gamma^2} |t|} \RR) ~U^\dag_{i k}~U_{k j} } \, . 
\ee
In the opposite under-damped regime ( $\gamma \ll \Omega_k$) the asymptotic expression for the propagators is:
\be
\label{Delta_under}
\Delta_{i j}(t) 	&=& \displaystyle{ \frac{e^{-\frac{\gamma}{2}~|t|}}{2 ~M^2~\gamma~\Omega_k^2} ~\cos \LL( \Omega_k |t| \RR) ~U^\dag_{i k}~U_{k j}} \, ,\\
\label{DeltaV_under}
\mathcal{V}_{i j}(t) 	&=& \displaystyle{ \frac{e^{-\frac{\gamma}{2}~|t|}}{2 ~M~\Omega_k} ~\sin \LL( \Omega_k |t| \RR) ~U^\dag_{i k}~U_{k j} } \, . 
\ee

Eq. (\ref{PIstep5}) is one of the central results of this work. It shows that the conditional probability $P({\bf k}_f, t| {\bf k}_i)$ for the dissipative dynamics of the quantum charge can be written in a form which is \emph{formally} analog to that of a vacuum-to-vacuum amplitude, in an effective zero-temperature  Dirac-like quantum field theory. This analogy is quite remarkable, since our theory describes an open system and is fully consistent with the fluctuation-dissipation relation. We also emphasize that the path integral expression (\ref{PIstep4}) is  real and positive definitive as it yields directly the conditional probability, even though it is \emph{formally} equivalent to probability amplitude (namely a two-point Green's function) in the effective quantum field theory.

We remark again that a particularly attractive feature of this formulation is that it does not involve the Keldysh contour. 
This strongly simplifies the formalism, since one does not need to introduce different types of Green's functions to distinguish between different sectors of the Keldysh contour. Instead, one can adopt  time-ordered Feynman propagators and apply text-book perturbative and non-perturbative quantum field theory techniques to evaluate directly the conditional probability and the expectation values of operators.

\section{Perturbation Theory and Feynman Diagrams}\label{Pert_Theory}\label{PTheory}
\begin{figure}[t!]
\begin{center}
\includegraphics[width=3cm]{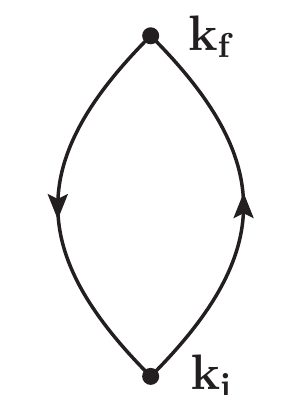}
\caption{The diagram corresponding to the unperturbed contribution to the conditional probability.}  
\label{diagrams_0}
\end{center}
\end{figure}
In the short-time  and  weak-coupling regimes, the conditional probability $P_t\LL(\kk_f|\kk_i\RR)$ can be computed analytically in  a perturbation theory derived   by performing a Taylor expansion of the exponents in Eq. (\ref{PIstep5})  in powers of the interaction terms
\bea
\label{intV1}
V_1 & = & -\frac{M \gamma}{\beta \hbar^2} ~\int_0^t d t' \int_0^t d t'' \bar{\psi}_{\bf m}(t') ~f^i_{\bf m n} ~\psi_{\bf n}(t') ~\Delta_{i j}(t'-t'')  ~\bar{\psi}_{\bf m'}(t'') ~f^j_{\bf m n}~\psi_{\bf n'}(t'') \, , \\
\label{intV2}
V_2  & = &  \frac{i}{4\hbar} ~ \int_0^t d t' \int_0^t d t'' \bar{\psi}_{\bf m}(t') ~\gamma_0 ~f^i_{\bf m n} ~\psi_{\bf n}(t') ~\mathcal{V}_{i j}(t'-t'') ~\bar{\psi}_{\bf m'}(t'') ~f^j_{\bf m n}~\psi_{\bf n'}(t'')\, .
\ee
The conditional probability is then written as
\be
\label{series}
P_t \LL(\mathbf{k_f} |\mathbf{k_i} \RR) = \sum_{i}^\infty P_t^{(i)}\LL(\mathbf{k_f} |\mathbf{k_i} \RR) \, , 
\ee
where $P_t^{(0)}\LL(\mathbf{k_f} |\mathbf{k_i} \RR)$ corresponds to the unperturbed conditional probability, which neglects all the couplings between the hole, the heat bath and the vibronic modes,
\be
\label{P0}
P_t^{(0)} \LL(\mathbf{k_f} |\mathbf{k_i} \RR) &=& \frac{-1}{Z^{(0)}} \int \mathcal{D}\bar{\psi}~ \D \psi~ e^{-\mathcal{L}_1(t,0)} ~ \left(\bar \psi_{\kk_f}(t) ~\gamma_- \gamma_5~  \psi_{\kk_f}(t)~ \bar{\psi}_{\kk_i}(0)~\gamma^+ \gamma_5~\psi_{\kk_i}(0)\right)~e^{-\frac{i}{\hbar}S_0[\bar\psi, \psi]} \, .
\ee
Its normalization factor $Z^{(0)}$ can be written in path integral form as:
\be\label{Z0}
Z^{(0)} = \int \mathcal{D}\bar{\psi}~ \D \psi~ e^{-\mathcal{L}_1(t,0)} ~ \sum_{\kk_f} \left(\bar \psi_{\kk_f}(t) ~\gamma_- \gamma_5~  \psi_{\kk_f}(t)~ \bar{\psi}_{\kk_i}(0)~\gamma^+ \gamma_5~\psi_{\kk_i}(0)\right)~e^{-\frac{i}{\hbar}S_0[\bar\psi, \psi]} \, .
\ee

 The leading-order perturbative correction in the series (\ref{series}) reads
\be\label{P1}
P_t^{(1)} \LL(\mathbf{k_f} |\mathbf{k_i} \RR) &=& \frac{-1}{Z^{(0)}+Z^{(1)}} \int \mathcal{D}\bar{\psi}~ \D \psi~ e^{-\mathcal{L}_1(t,0)} ~ \left(\bar \psi_{\kk_f}(t) ~\gamma_- \gamma_5~  \psi_{\kk_f}(t)~ \bar{\psi}_{\kk_i}(0)~\gamma^+ \gamma_5~\psi_{\kk_i}(0)\right)~(V_1+ V_2)~e^{-\frac{i}{\hbar}S_0[\bar\psi, \psi]},\qquad
\ee
where the corresponding leading-order correction to the normalization factor is 
\be\label{Z1}
Z^{(1)} &=&\int \mathcal{D}\bar{\psi}~ \D \psi~ e^{-\mathcal{L}_1(t,0)} ~ \sum_{\kk_f} \left(\bar \psi_{\kk_f}(t) ~\gamma_- \gamma_5~  \psi_{\kk_f}(t)~ \bar{\psi}_{\kk_i}(0)~\gamma^+ \gamma_5~\psi_{\kk_i}(0)\right)~(V_1+V_2)~e^{-\frac{i}{\hbar}S_0[\bar\psi, \psi]} \, .
\ee

 Eq.s (\ref{P0}), (\ref{Z0}), (\ref{P1}) and (\ref{Z1}) correspond to  correlation functions in the free limit for the effective Dirac-like quantum field theory. 
According to Wick's theorem, these Green's functions can be evaluated
by considering the sum of all possible contractions
between the $\psi$ and $\bar \psi$ fields and replacing each contraction with time-ordered Feynman propagator:
\be 
\contraction{}{\psi}{{}_i(t'')\>{}}{\bar \psi}  \psi_i (t'')\>{}\bar \psi_j (t')\rightarrow G_{\bf ij}^0 (t''-t')= 
 ~ \displaystyle{ V^{\dagger}_{\mathbf{is}}~  e^{ -  \frac{i}{ \hbar}~f^0_{\bf s} (t'' - t')}~V_{\mathbf{sj}} }~ \LL[ ~\gamma_+ \theta(t''-t') - \gamma_- \theta(t' - t'') ~\RR] \, ,
\ee
where the matrix elements $V_{\bf ij}$ define the unitary transformation which diagonalizes the hopping matrix $f^0_{\bf ij}$, while $f^0_{\bf s}$ are the corresponding eigenvalues. 
\begin{figure}[t!]
\begin{center}
\includegraphics[width=18cm]{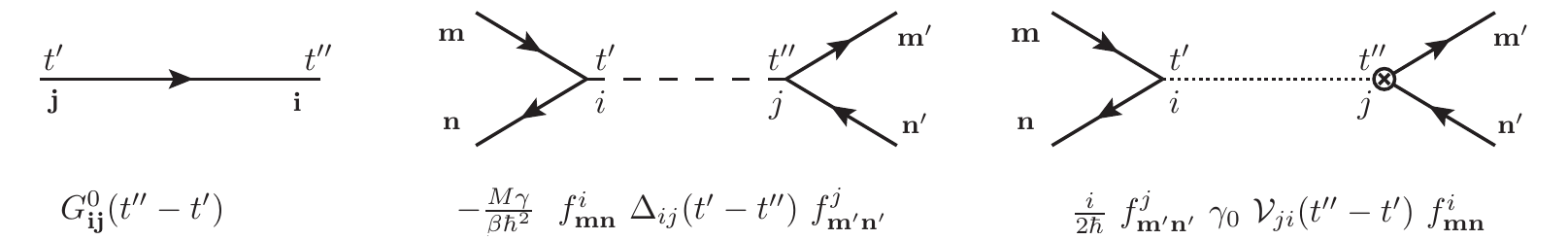}
\caption{Feynman rules for the effective field theory for charge propagation in the macromolecule. On the left panel, we show the hole's Feynman propagator, on the center panel the effective interaction  $V_1$, and on the right panel the effective interaction $V_2$.}
\label{FeynmanRules}
\end{center}
\end{figure}

The  Wick contractions are most conveniently defined and computed using a diagrammatic technique, i.e. by applying the Feynman rules shown in Fig.~\ref{FeynmanRules}. Just like in the standard quantum field theory, one can prove that the corrections to the normalization factor $Z^{(1)} $ exactly cancel out  with the contribution of disconnected diagrams, order-by-order in perturbation theory.

From the zero-th order diagram shown in Fig.~\ref{diagrams_0}  we readily re-obtain the  unperturbed conditional probability $ P^0_t(\kk_f | \kk_i)$ 
\be\label{P0result}
 P^{(0)}_t(\kk_f | \kk_i)  = ~ -G_{\bf k_i k_f}^0 (-t) ~G_{\bf k_f k_i}^0 (t) = ~V^{\dagger}_{\mathbf{k_i n}} ~ e^{ i ~f^0_{\bf n}/ \hbar \,t} ~V_{\mathbf{n k_f}} \, V^{\dagger}_{\mathbf{k_f s}} ~ e^{- i ~f^0_s / \hbar  \,t} V_{\mathbf{s k_i}} ~ = ~ \LL| G_{\kk_f \kk_i}^0 (t) \RR|^2 \, .
\ee

\begin{figure}[t!]
\begin{center}
\includegraphics[width=12cm]{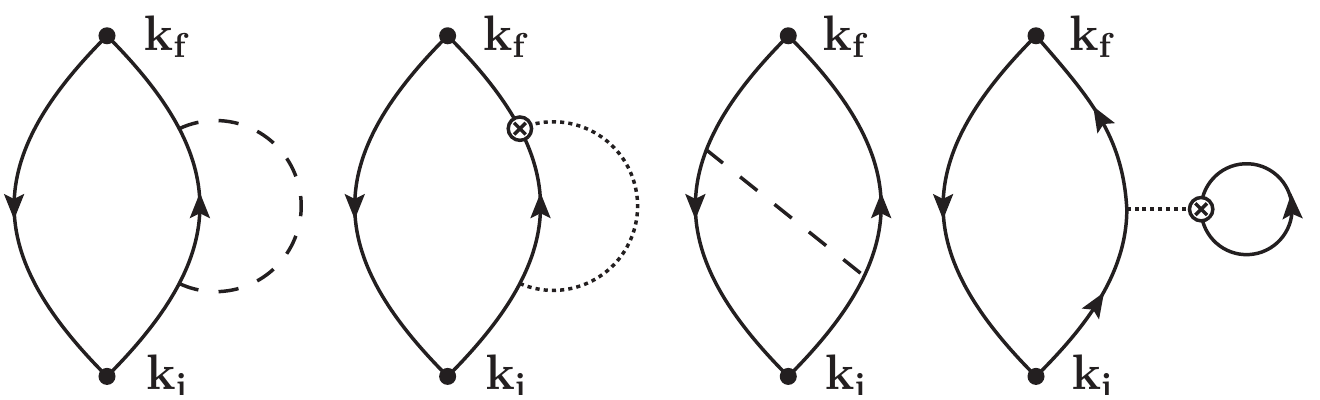}
\caption{The  diagrams involved in the leading-order correction to the conditional probability. 
The first  two diagrams from the left-hand-side are called ``self-energy''-type. The third diagram is called ``crossing''-type and the last is a ``tad-pole'' diagram. Analog diagrams exist for ``self-energy'' and ``tad-pole'' diagrams, in which the vibron propagators are coupled to the hole propagator on the left. }  
\label{diagrams_1}
\end{center}
\end{figure}

The different types\footnote{Clearly, in addition to the  diagrams shown in Fig.~\ref{diagrams_1}, there are also equivalent ones in which the vibronic propagators are emitted and absorbed by the backward propagating fields.} of diagrams which contribute to  $ P^1_t(\kk_f | \kk_i)$  are shown in Fig.~\ref{diagrams_1}.  We note that the first two of such diagrams contain a ``self-energy''-type correction to one of the propagators. The third diagram contains the interaction of forward and backward propagating holes and will be called the ``crossing''-type diagram.  Finally, the last diagram is a ``tad-pole''. 
After collecting all terms, we obtain the following expression for the first-oder correction to the conditional probability: 
\bea\label{P1corr}
P_t^{(1)} \LL(\mathbf{k_f} |\mathbf{k_i} \RR)  
 & = &  \frac{4 M \gamma}{\beta \hbar^2} 	~\text{Re} \bigg[ ~\int_0^t ~ d\tau d\tau' ~G^{0}_{\bf k_i k_f} (-t) ~G^{0}_{\bf k_f q'} (t - \tau') 
 ~f^{j}_{\bf q' s'} ~\Delta_{ji} (\tau'-\tau)     ~G^{0}_{\bf s's} (\tau'-\tau) ~f^{i}_{\bf sq} ~G^{0}_{\bf q k_i} (\tau) \bigg]  \nn \\
 & + &  \quad \frac{2}{ \hbar} 	\, 			~\text{Im} \bigg[ ~\int_0^t ~ d\tau d\tau' ~G^{0}_{\bf k_i k_f} (-t) ~G^{0}_{\bf k_f q'} (t - \tau') 
 ~f^{j}_{\bf q' s'} ~\mathcal{V}_{ji}(\tau'-\tau) ~G^{0}_{\bf s's} (\tau'-\tau) ~f^{i}_{\bf sq} ~G^{0}_{\bf q k_i} (\tau) \bigg]  \nn \\
 & +  & \quad \frac{2}{\hbar}	\,		 	~\text{Im} \bigg[ ~\int_0^t ~ d\tau d\tau' ~G^{0}_{\bf k_i k_f} (-t) ~G^{0}_{\bf k_f q } (t - \tau)
 ~f^{j}_{\bf q s}~G^{0}_{\bf s k_i} (\tau) ~\mathcal{V}_{ij}(\tau' - \tau) ~f^{i}_{\bf s's'}  \bigg]  \nn \\
 & + & \, \, \frac{2 M \gamma}{\beta \hbar^2} \,		~\int_0^t ~ d\tau d\tau' ~ G^{0 }_{\bf k_i s'}(t-\tau') ~f^{j}_{\bf s' q'} ~G^{0}_{\bf q' k_f}(\tau') 
 ~\Delta_{j i} \LL( \tau'-\tau \RR) ~G^{0}_{\bf k_f s}(t-\tau) ~f^{i}_{\bf sq} ~G^{0}_{\bf q k_i} (\tau) \, .
\eea
The first two lines are the contributions due to the ``self-energy"-type diagrams, the third is derived from the ``tad-pole'' diagram, and the last line is derived from the ``crossing''-type diagram. Further the details of this calculation are provided in  Appendix \ref{AppA}.

We conclude this section by discussing the regimes in which we expect the perturbative expansion to be applicable.  
 To this end, we introduce the explicit expressions for $G_{\bf ij}^0$, $\Delta_{ij}$, and $\mathcal{V}_{ij}$ into Eq. (\ref{P1corr}),  take the short time limit $t\ll 1/\Omega_k$, $t\ll 1/\gamma$,  and consider for simplicity only the over-damped and under-damped asymptotic expressions for the Green's functions. 

In the  in the over-damped regime, we find the conditions of validity
\bea
\frac{2 ~f^{i\, 2}_{\bf qs}~ t^2}{\beta ~M ~\Omega^2_k ~\hbar^2} ~&\ll&~ 1\qquad (\text{from the}~ V_1\text{-type interaction}) \, ,  \\ 
\frac{4 ~f^{i\, 2}_{\bf qs}~ t^2}{M ~ \gamma ~\hbar} ~~&\ll&~ 1\qquad (\text{from the}~V_2\text{-type interaction}) \, . 
\eea
In the  under-damped regime, the conditions of validity of the perturbative expansion are
\bea
\frac{2 ~f^{i\, 2}_{\bf qs}~ t^2}{\beta ~M ~\Omega^2_k ~\hbar^2} ~&\ll&~ 1 \qquad (\text{from the}~ V_1\text{-type interaction}) \, ,  \\ 
\frac{2 ~f^{i\, 2}_{\bf qs}~ t^2}{M ~ \Omega_k ~\hbar} ~&\ll&~ 1 \qquad (\text{from the}~V_2\text{-type interaction}) \, . 
\eea

We note that in both the under-damped and over-damped regimes  the perturbative approach is only valid at short times. This is completely expected, because the long-time and long-distance propagation necessarily involves multiple scattering of the hole with the heat bath and molecular vibrations, hence require a non-perturbative treatment.
By plugging order of magnitude estimates of the normal-mode frequencies 
($\Omega_k\sim ~ 10^{-3} ~ \text{fs}^{-1}$), the gradient of the hopping matrix elements ($f^i_0\sim 10^{-2}$ eV \AA$^{-1}$), and the viscosity ($\gamma=0.1  ~ \text{fs}^{-1}$) we find that, at room-temperature and in the over-damped limit, the $V_1$-type interaction is several orders of magnitude larger than the $V_2$-type, hence determines the range of validity of the perturbative expansion.  On the other hand, in the opposite under-damped limit, the driving interaction is $V_2$-type.

\section{Charge Propagation through a poly(3-alkylthiophene) Molecule}\label{model}
\begin{figure}[t!]
\begin{center}
\includegraphics[width=8cm]{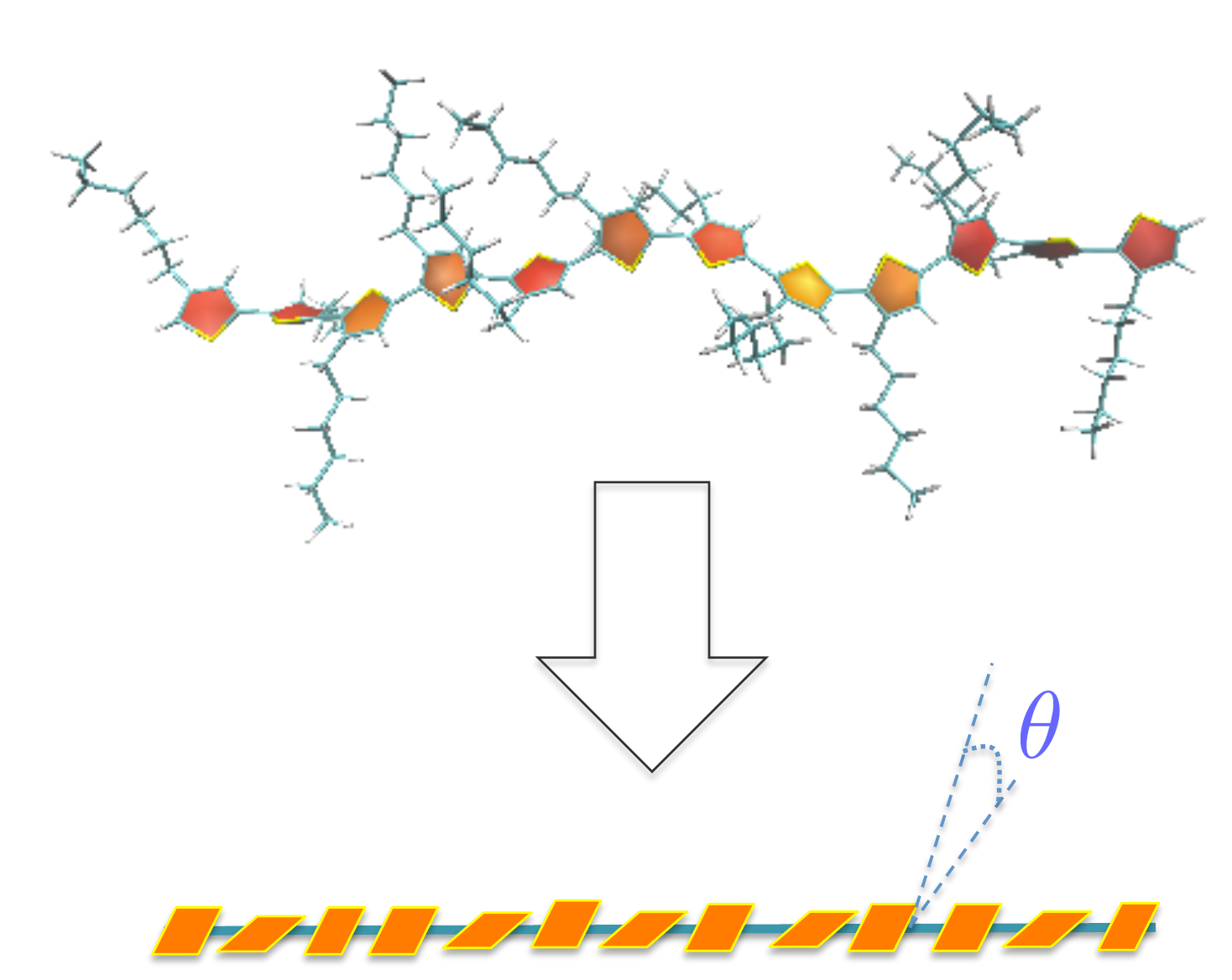}\caption{Upper panel:  three-dimensional structure of a P3HT polymer. Lower panel: the coarse-grained  representation corresponding to our effective model. }
\label{toy model}
\end{center}
\end{figure}

Let us now illustrate the formalism developed in the previous sections by investigating the intra-chain propagation of electron holes through the backbone of a  poly(3-alkylthiophene) (P3HT) polymer. Quasi-cristalline materials made of inter-digited PH3T polymers have received much attention, in connection with the possibility of realizing nano-scale organic transistors~\cite{PH3Trev, organic3,organictransistor}. 
The atomistic three-dimensional structure of a PH3T molecule is  shown in the upper panel of Fig.~\ref{toy model}. 

Here, we are only interested in providing  a qualitative description of the charge propagation in such systems, leaving a more sophisticated quantitative description to our future work. Our main goal is to  estimate the order-of-magnitude of the range of times and distances over which the perturbative approach is applicable. Secondly, we are interested in comparing the probability densities obtained by means of the perturbative calculation and by brute-force integration of the equation of motion (\ref{EoMs}). Finally, we investigate how the different charge-charge effective interactions which appear in Eq. (\ref{PIstep4}) affect the charge dynamics and in particular  quantum de-coherence and re-coherence phenomena. 

\subsection{Coarse-Grained Model}

In order to address these points it is sufficient to adopt a simple coarse-grained representation of the molecule, in which side-chain degrees of freedom are not taken explicitly into account. Furthermore, the molecular potential energy function is assumed to effectively depend only of the dihedral angles formed by neighboring aromatic rings. Hence, the molecular configuration is specified by the set of dihedral angles  $\Theta=(\theta_1, \ldots, \theta_N)$
and the chain is mapped into an effective one-dimensional system consisting of $N$ plaquettes which can rotate around their symmetry axis, as  sketched in the lower-right panel of Fig.~\ref{toy model}.

The potential energy of a molecular configuration is approximated with sum of pairwise terms, each of which depends on the relative dihedral angle of two consecutive monomers,
\be
U(\Theta)= \sum_{i} u(\theta_i- \theta_{i+1}) \, .
\ee

\begin{figure}[t!]
\begin{center}
\includegraphics[width=7cm]{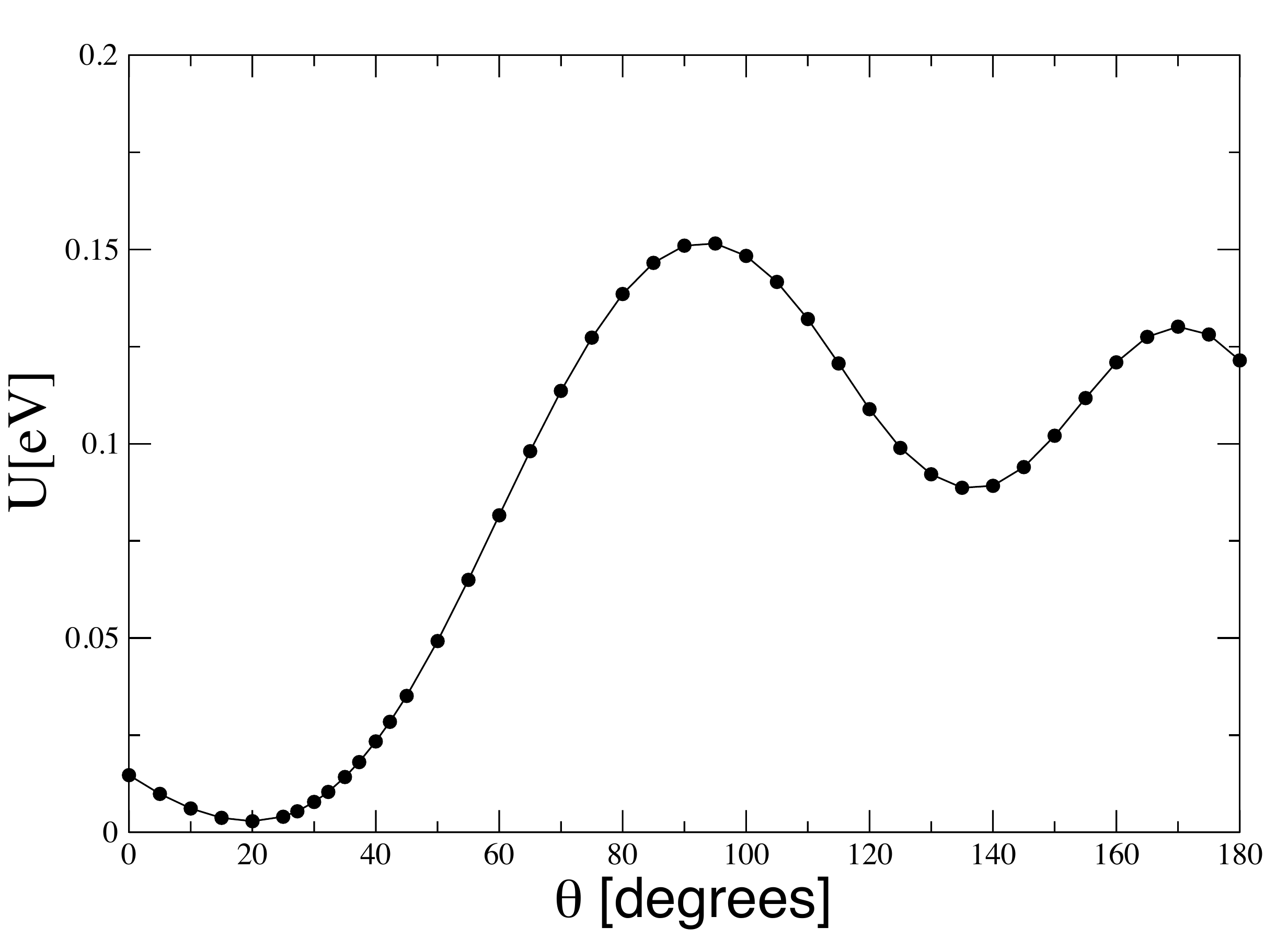}
\includegraphics[width=7cm]{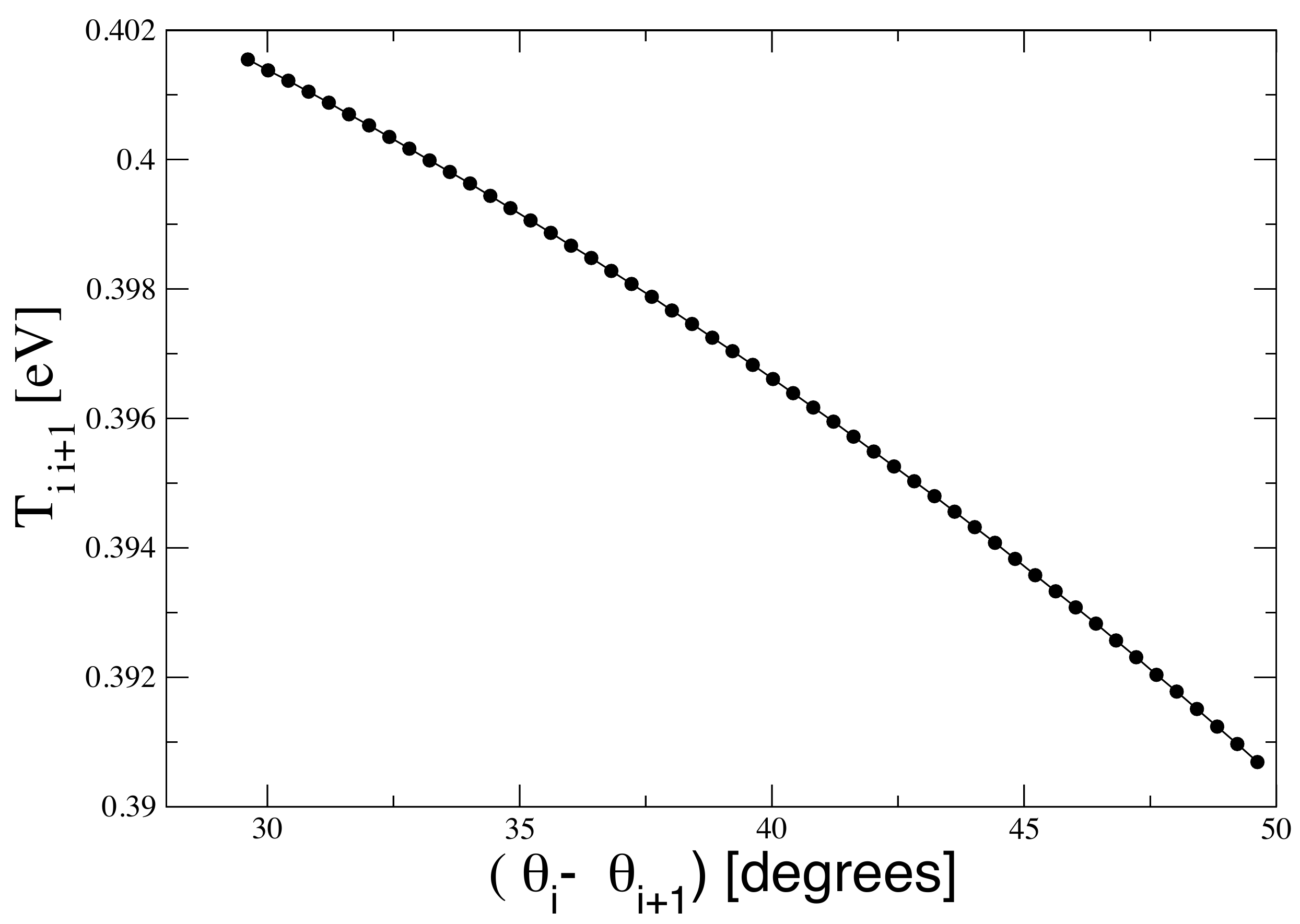}\caption{Left panel: DTP-B calculation of the molecular potential energy of two consecutive monomers as a function of the difference of dihedral angles. The mechanical equilibrium is retained at an angle $s^i$ which in this case is  $\simeq 20^o$. In general $s^i= + (-) \theta_0$ for $i$ odd (even), with $\theta_0=20^o$.  Right panel: DFT-TB calculation of the transfer integrals $T_{i+1, i}$ as a function of the difference in the dihedral angles $(\theta_{i+1}-\theta_i)$, in the proximity of the mechanical equilibrium configuretion $\theta_0$.  }
\label{DFTB}
\end{center}
\end{figure}
We have obtained the pair-wise interaction energy  $u(\theta_i- \theta_{i+1})$ as a function of the relative angle $\theta = \theta_i- \theta_{i+1}$ from  DFT-TB electronic structure calculations, using the \emph{DFTB+} package~\cite{DFTBrev}. The results are shown in the left panel of Fig.~\ref{DFTB}.
In the lowest-energy configuration, the aromatic rings in the different residues form a relative dihedral angle of $(-1)^i~ \theta_0$ (where $\theta_0=20^o$ and $i$ is the monomer index).

The conformational dynamics is therefore described by the Hamiltonian:
\be
H_M= \frac{1}{2 I}~~ \sum_{i=1}^{N} p_i^2 + U(\Theta) \, ,
\ee
where  $I$ is the momentum of inertia of the monomers (including the contribution from the atoms in the side-chain) and $p_i= I \dot \theta_i$ is the canonical momentum conjugated to the $\theta_i$ generalized coordinate.

At room temperature, this system performs only small thermal oscillations around the minimum-energy configuration. Hence, we can perform a small-angle expansion, leading to the simple harmonic form:
\be
H_M \simeq \frac{1}{2I}~ \sum_{i=1}^{N}p_i^2 + \sum_{i=1}^{N-1} \frac{\kappa}{2} (\theta_{i+1}-\theta_i + (-1)^i \theta_0)^2 + \frac{\kappa}{2} \theta_1^2 + \frac{\kappa}{2} \theta_N^2 \, .
\ee
The last two terms follow from assuming that the first and last monomers in the chain are  bond to external non-conducting leads which tend to align them horizontally.  

The momentum of inertia $I$ can be calculated directly from the three-dimensional structure of the chain  and 
affects the frequencies of the chain's normal modes of oscillations,   
\be
\omega^2\sim \frac{\kappa}{I} \, .  
\ee
In the systems which are of technological and experimental interest,  P3HT polymers are embedded in organic frameworks. In such a configuration, the chain exchanges energy with neighboring molecules, which play the role of a heat bath. In addition, the steric interaction with neighbors generates strong constraints on the chain dynamics and in particular affects the amplitude and frequencies of thermal oscillations. 
In order to account for this effect, we consider an effective model in which the spring constant $\kappa$ which appears in the
molecular potential energy function $U(\Theta)$ is artificially rescaled in such a way that the typical square fluctuations of the dihedral angles around their equilibrium values, $ \langle \Delta \theta^2 \rangle_{MD}= \frac{1}{N} \sum_i \langle ((\theta_{i+1}-\theta_i + (-1)^i~\theta_0)^2\rangle_{MD}$ matches the value obtained from classical molecular dynamics simulations for a system of inter-digited PH3T polymers~\cite{tesi}: 
\be
\kappa\rightarrow \kappa_{eff} \, , \qquad \text{and} \qquad \langle \Delta \theta^2 \rangle_{MD} \simeq \frac{k_B~T}{\kappa_{eff}} \, .
 \ee 
 
 Also the hopping matrix elements $T_{i i+1}$ between neighboring monomers and the on-site energies $e_i$ as a function of the chain's configurations have been obtained by DFT-TB calculations. In analogy with molecular energy calculations, the  transition matrix elements  $T_{ii+1}$ have been computed assuming that they effectively depend only on the relative angles, $ \LL|\theta_{i+1}-\theta_i \RR| $. For sake of simplicity, we have taken the on-site energies $\epsilon_i$ to be constant and equal to the value at the mechanical equilibrium configuration. This choice can be motivated by the observation made by different groups (see e.g. Ref.~\cite{DNA_theory2}) that fluctuations of the on-site energies have a much smaller effect on the electric conduction than fluctuations of the transfer matrix elements.  The results for $T_{i i+1}(\theta_{i+1}-\theta_i)$, in the vicinity of the equilibrium configuration $\theta_0$ are reported in the right panel of Fig.~\ref{DFTB}.
By assuming a linear approximation, Eq. (\ref{HOPPING_matrix}) takes the form
\be
f_{m n}\LL(\theta_i\RR) & = & f^{0}_{mn} + f^{1}_{mn}~\LL(\, \LL| \theta_{m} - \theta_{n} \RR| - \theta_0 \RR) \, ,
\ee
where
\bea
\label{hop_matrix}
f^{0}_{mn} & = & T_0~ \LL( 1-\delta_{m n} \RR) - e_0 ~\delta_{m n} \\
f^{1}_{mn} & = & T_1~ \LL( 1- \delta_{mn} \RR)\, .
\eea 
Finally,  the viscosity parameter $\gamma$ may be determined from MD simulations by computing the velocity autocorrelation function. On the other hand, we have observed that the results of the perturbative calculation depend very weakly on the this parameter. Hence, for sake of simplicity, here we assume a reasonable value $\gamma = 0.1$~fs$^{-1}$. 

The numerical values of the parameters of this coarse grained model are summarized in Table \ref{TABparam}. 
\begin{table}[t!]
 \centering
\begin{tabular}{|c|c|c|c|c|c|c|c|c|c|}
\hline
  $~e_0~\LL[eV\RR]~$   & $~T_0~\LL[eV\RR]~$ & $~T_{1}~\LL[eV\RR]~$ & $~\theta_0~\LL[deg\RR]~$ &  $~\gamma~\LL[fs^{-1}\RR]~$ & $~\kappa~\LL[eV\RR]~$ & $\kappa_{eff}~\LL[eV\RR]~$ & $~T~\LL[\mbox{}^o K\RR]$ & $~I~\LL[uma ~\AA{}^{2^{}}\RR]~$ \\ \hline 
 5.4  & 0.4   &  0.06   &    20  &  0.1 &  0.13 & 0.20 & 300 & 3400\\ \hline
\end{tabular}
\caption{Parameters or the coarse-grained model which describes intra-chain hole propagation in P3HT polymers. The PH3T chain investigated in the simulations consists of 9 monomers.}
\label{TABparam}
\end{table}
The equilibrium configuration can be chosen to be:
\be
\theta_i = \LL\{ \begin{array}{cc}
0 & \text{if $i$ odd,} \\ 
\theta_0 &  \text{if $i$ even.} 
\end{array} \RR.
\ee
The hole propagator $G^0_{l m}(t)$ is constructed  by diagonalizing the $\hat f^0$ matrix, defined in Eq.~(\ref{hop_matrix}). Its $n$-th eigenvector reads
\be
\phi_n(j)= \sqrt{\frac{2}{N+1}}~\sin\left[\frac{n~\pi}{(N+1)}~j\right] \, . 
\ee
The corresponding eigenvalue is
\be
E_n = - e_0 - 2 T_0~\cos(k_n ) \, ,
\ee
where $k_n$ is the wave-vector
\be
\label{wavenumbers}
k_n = \frac{\pi n}{(N+1)},\quad n=1,2,\ldots, N  \, .
\ee
Hence, the hole Feynman  propagator is given by
\be
G^0_{j_f, j_i}(t)
&=& \sum_{n=1}^{2(N+1)}~\phi^\ast_n(j_f) ~\phi_n(j_i)~e^{-\frac{i}{\hbar} E_n t}~\left( \gamma^+  \theta(t) -   \gamma_- \theta(-t) \right).
\ee

With the present set of model parameters, we find that the friction coefficient $\gamma$  is significantly larger than the typical frequencies of normal models, i.e. $\gamma \gg \Omega_k$. Hence,  we can use the over-damped limit for the vibronic propagators.  

\subsection{Time Evolution of the Conditional Probability}

\begin{figure}[t!]
\begin{center}
\includegraphics[width=10cm]{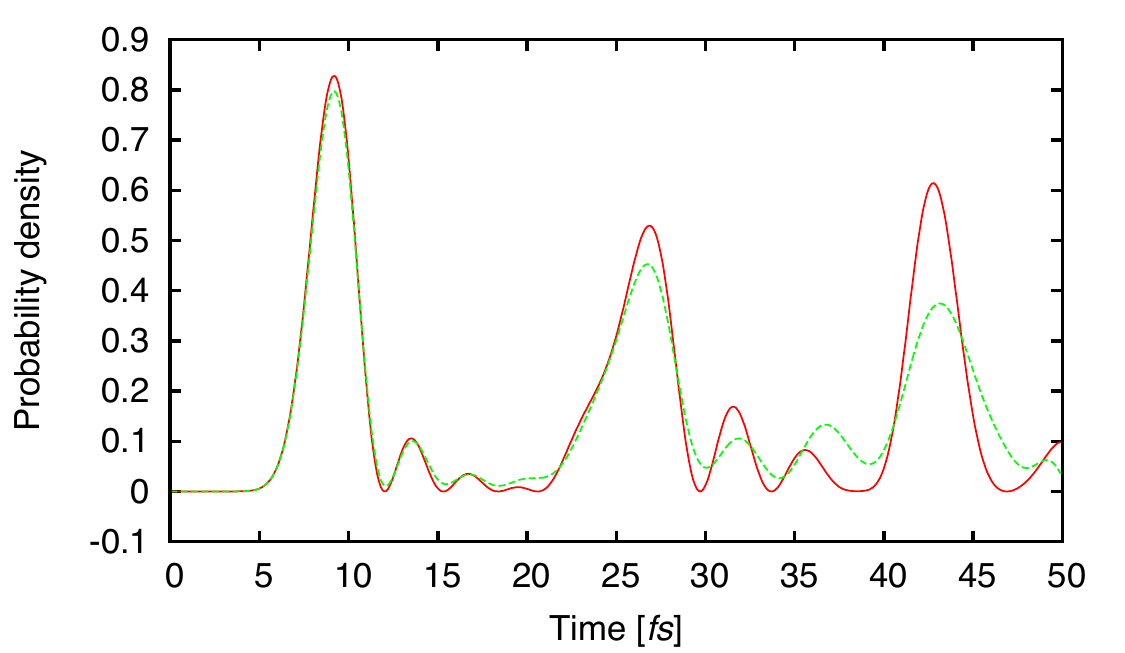}
\caption{ Time evolution of the charge density at the right endpoint of the chain, assuming that a hole is initially created at left-end of the chain. The solid line represents the unperturbed prediction, the dashed line includes the effects of the coupling with the molecular vibration and the heat bath to 
leading-order in perturbation theory. }
\label{endpointcharge}
\end{center}
\end{figure}

We assume that the hole is initially created at the left end of the chain and evolves in time at a temperature of 300~K. In Fig.~\ref{endpointcharge} we show the results of our leading-order perturbative calculation of the hole density at the opposite end of the chain as a function of the time interval $t$, i.e. $P_t(\kk_N|\kk_1)$, where  $N=9$. 

Some comment on these results are in order. First of all, we notice the existence of three peaks in the charge density, at times $t\sim 10, 20, 40$~ps. These corresponds to integer multiples of the time interval it takes the hole to run along the entire chain.  
Next, we note that the scattering of the holes with the molecular normal modes and with the heat bath slows down the charge propagation, as expected. This is evident from the fact that  the probability of observing the hole at the right end-point of the chain as a function of time is reduced once the perturbative corrections are included. Correspondingly, the times at which the hole rebounds to the right end-point of the chain are delayed by the interactions. We recall that the norm of the conditional probability is conserved up to corrections which are of higher-order in the perturbative expansion. Hence, the reduction of the charge density at the end-point of the chain implies that the scattering distributes the charge density in the central region of the chain.    

We observe that the correction to the conditional probability starts to be of the same order of the unperturbed prediction starting from time intervals of the order of $\sim 40$~fs.  Beyond this time scale, the perturbative approach breaks down and one has to resort on non-perturbative approaches in which many Feynman diagrams are 
re-summed.

\subsection{Quantifying the Loss of Quantum Coherence}

We have seen that analytic perturbative calculations break down beyond a few tens of fs, hence do not represent a useful tool to investigate the long-time long-distance dynamics of hole propagation. On the other hand, they provide a valuable tool to gain analytic insight into the physical mechanisms which drive de-coherence and re-coherence during hole propagation across the chain. 

As measure of the degree of quantum coherence in the dynamics of an open system, we consider the ratio~\cite{coherence1, coherence2}:
\be
R(t) = \frac{\text{Tr}\rho^2(t)}{\text{Tr}\rho(t)} \, .
\ee   
In Appendix \ref{quantumdecoherence} we show that this ratio is identically equal 1 for pure states (corresponding to fully coherent propagation), and that it is smaller than 1 for mixed states.
  
\begin{figure}[t!]
\begin{center}
\includegraphics[width=10cm]{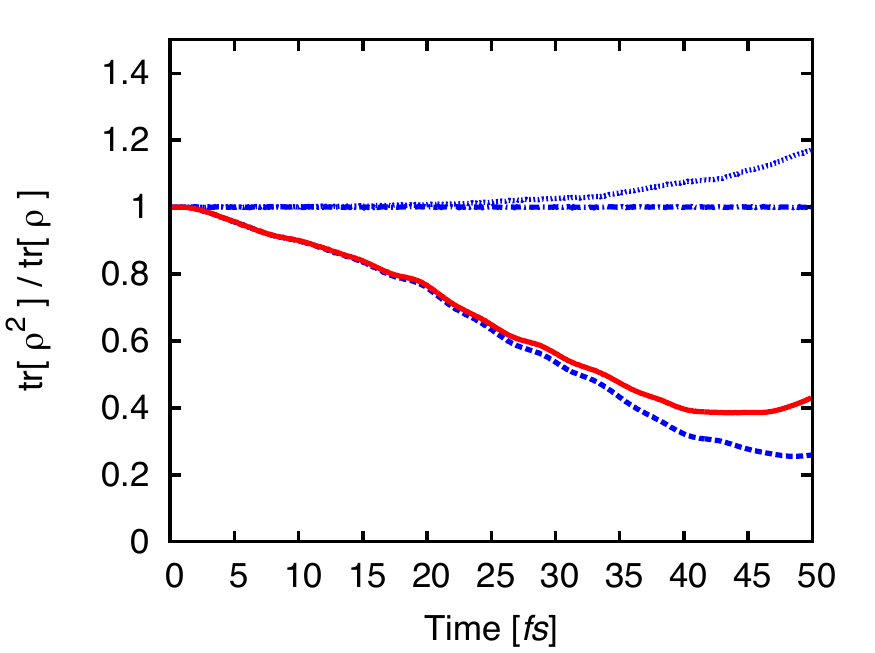}
\caption{Perturbative calculation of the different contribution to the ratio $R(t)$ which quantifies the de-coherence effects in hole propagation. The dot-dashed line is the unperturbed result, which remains coherent at all times. The solid line is the result including the leading order correction. Dotted and dashed lines correspond to the contribution of ``self-energy"-type and ``crossing"-type, respectively. For this model, the ``tad-pole"-type contribution is identically zero. }
\label{decoherence}
\end{center}
\end{figure}

In Fig.~\ref{decoherence} we compare this ratio for the model under consideration in the limit of unperturbed propagation and including the leading-order scattering with the molecular vibrations and with the heat bath. We see that the interaction with the environment suppresses the quantum coherence on time scales which are of the order of 10 fs. 

It is interesting to compare the contribution to $R(t)$ coming from the different Feynman diagrams shown in Fig.~\ref{diagrams_1}. We find that the quantum decoherence is driven by the ``cross"-type diagram shown in figure \ref{diagrams_1}, which tends to correlate the field components associated to  propagation forward and backwards in time.
In the equivalent zero-temperature quantum-field theory effective picture, the quantum decoherence emerges as a result of the formation of a particle-antiparticle ``bound state". On the other hand, the so-called ``self"-type diagrams act in the opposite direction, slowing down the overall rate of quantum decoherence. 

The identification on the diagram which drives the quenching of $R(t)$ with time offers a scheme to study how the chemical and mechanical properties of the macromolecule affect the quantum decoherence of the propagating excitation. Indeed, by varying the parameters of the effective theory (namely the spectrum of normal modes $\omega_k$ entering in the vibron propagators, and unperturbed tight-binding matrix elements $f^0_{\bf lm}$ and their gradient $f^i_{\bf lm}$, which enter  the effective interaction vertexes)  and  computing the corresponding relative weight of the ``cross"-type diagram,  one may in principle identify what properties of the macromolecular system are most effective in suppressing (or enhancing) the  quantum coherent transport. This information may be useful e.g. in the context of the study of exciton propagation in photosynthetic complexes, which have been found to display quantum coherent dynamics over surprisingly long time intervals.

\subsection{Comparison between the perturbative estimate and the result of integrating the quantum/stochastic equations of motion}

The perturbative approach developed in the previous sections allows to analytically compute the charge density, in the range of time intervals $0< t \lesssim50$~fs.
It is interesting to compare the perturbative calculation with the results of non-perturbative numerical simulations, obtained by averaging over many independent solutions of the set of quantum/stochastic equations of motion defined in Eq.~(\ref{EoMs}) and derived in Ref.~\cite{boninsegna}.   
On the one hand, this provides a non-trivial test for the perturbative scheme developed in this work. On the other hand, it offers an estimate of the statistical accuracy which is needed in order to resolve the effects of the interactions on the charge propagation dynamics in non-perturbative numerical simulations. 

\begin{figure}[t!]
\begin{center}
\includegraphics[width=12cm]{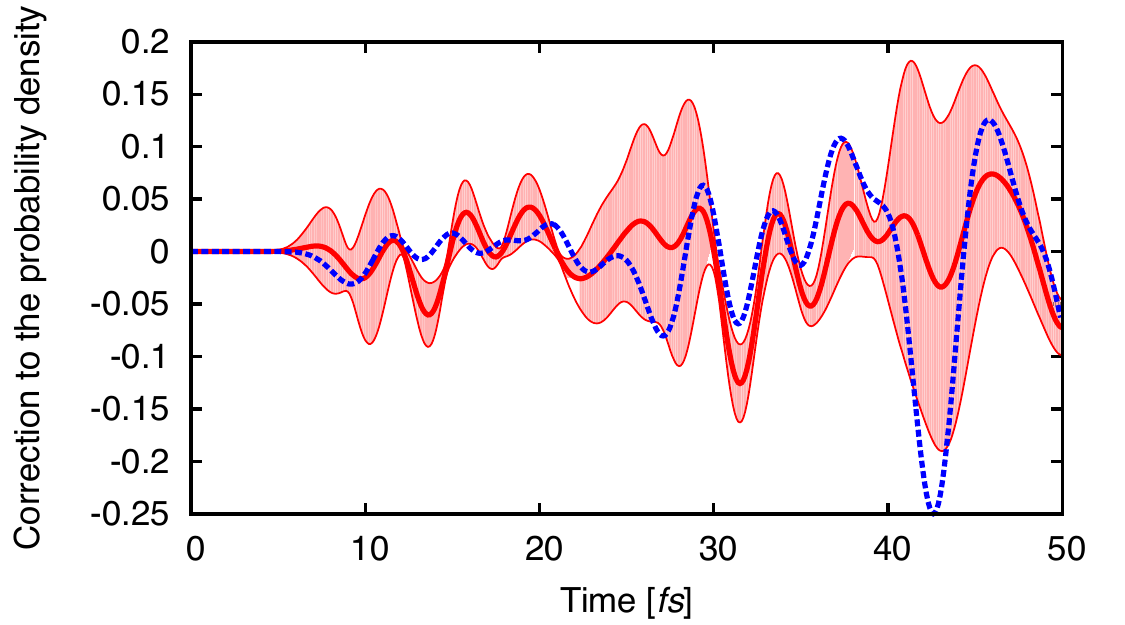}
\caption{ Time evolution of the charge probability density at the right end-point of the chain assuming that the  hole is initially created at left-end of the chain. We compare  non-perturbative numerical calculation obtained by integrating the Eq.s (\ref{EoMs}) (solid line) and the analytic perturbative calculations (dotted line). The shaded area represents the statistical uncertainty on the non-perturbative calculation. }
\label{comparison}
\end{center}
\end{figure}

In Fig.~\ref{comparison} we present the  difference  between the interacting and the free conditional probabilities 
$P_{t}(N, 1)-P^{(0)}_{t}(N, 1)$, evaluated in the perturbative and non-perturbative methods (we recall that $N=9$). 
The shaded area represents the statistical error in numerical simulations, which is estimated from the variance calculated from 10000 independent trajectories.  Accumulating this statistics required about 6 Central Processor Units (CPU) hours of simulation on a regular desktop. By contrast, the perturbative estimates took about a minute on the same machine. 

 We find that the two approaches are quantitatively consistent with one another, even at time scales of the order of 50~fs. Beyond such a time scale the perturbative approach becomes unreliable and the comparison is meaningless.  
 
It is important to emphasize that these two methods are based on different approximations. In particular, the algorithm defined by Eq.s~(\ref{EoMs}) was obtained by neglecting the fluctuations of the coherent fields 
around their functional saddle-point solution. At such a saddle-point, the forward- and backward- propagating fields are identical,  $\phi'(t)=\phi''(t)$. 
The leading-order perturbative estimate goes beyond such a  saddle-point condition and accounts for independent quadratic fluctuations on $\phi'(t)$ and on
$\phi''(t)$.
The relatively good agreement between the two calculation schemes at short times can be used as an argument in favor of the accuracy of the saddle-point approximation used in the non-perturbative approach.

\section{Conclusion and Outlook}
\label{conclusions}

In this work we have introduced a path integral approach to investigate the real-time dynamics of a quantum excitation propagating through a macromolecular system. In our starting model, all the atomic coordinates are explicitly taken into account, the dynamics of the quantum excitation propagating throughout the molecule is described by an effective configuration-dependent tight-binding Hamiltonian, and the heat bath is represented by the Caldeira-Leggett model. 

By adopting the first quantization for the molecular and the heat bath variables, and the second-quantization for the quantum excitation variables, we have been able to analytically trace out from the density matrix both the heat bath and the atomic nuclear dynamics. The result is a path integral formulated solely in terms of the quantum excitation's degrees of freedom. 
In contrast to other  field-theoretic approaches to charge transport in macromolecules~\cite{orland1, orland2}, in the present effective theory approach fluctuation-dissipation effects are fully taken into account. In addition, the free parameters can be obtained directly from microscopic quantum chemistry calculations. 

A first important advantage of our effective field theory formalism is that it makes it possible to describe real-time dynamics in an open quantum system without employing the Keldysh contour: indeed, the degrees of freedom associated to the backwards time-propagation in the Keldysh formalism can be effectively re-interpreted as anti-matter components of forward propagating quantum fields. In this way, the path integral which yields  the reduced density matrix for the quantum excitation becomes \emph{formally} equivalent to one associated to a standard vacuum-to-vacuum two-point correlation function, in a closed system at zero-temperature. This formal analogy is quite useful, as it immediately yields the  Feynman rules to compute by perturbation theory the corrections to the density matrix due to the interaction between the propagating excitation, the atomic coordinates and the heat bath degrees of freedom. 

The possibility of performing analytic calculations in the weak-coupling and short-time regime opens the door to a detailed investigation of the effects which generate quantum decoherence in molecular systems coupled to a heat bath. We have identified a specific Feynman diagram which dominates the dissipation of the quantum coherence, and correlates forward and backward propagating fields. 
In the future, it  would be interesting to perform the same calculation for a class of simple models, in order to clarify whether the dominance of this diagram is a universal property of all open quantum systems, and if so unveil its physical interpretation. 

For illustration purposes, we developed a coarse-grained model and applied this formalism to investigating the intra-chain propagation of holes in a P3HT polymer. We found that the propagation can be described in perturbation theory up to about 40-50 fs. 

Beyond that time scale, non-perturbative approaches are required. Since the coherent-state path integral is affected by a dynamical sign problem, Monte Carlo approaches would be challenging.  An alternative non-perturbative approach consists in directly integrating the quantum and stochastic equation of motions (\ref{EoMs}) which follow from a functional saddle-point approximation. The comparison with the analytic perturbative calculations has shown that the underlying saddle-point approximation is quite robust. However, a large statistics seems to be necessary in order to resolve the tiny effects of the interaction with the heat bath and with the vibronic modes.   
An alternative non-perturbative approach which would not involve stochastic averages and MD simulations could be provided by the  self-consistent saddle-point approximation of Eq. (\ref{PIstep4}). We plan to develop and test such an approach in our future work. 

Finally, an obvious direction for our future research consists in developing the formalism to compute the electric current induced by an external electric field. 
   
\section*{Acknowledgements}
 
 We thank G. Lattanzi and D. Alberga suggesting to investigate charge propagation through P3HT polymers. We also acknowledge an interesting discussion 
with Prof. Marco Garavelli. We also thank Prof. Marcus Elstner of Karlsruhe Institute of Technology for providing parameters files for \emph{DFTB+}. 

\appendix

\section{Details on the vibronic Green's functions structure and the perturbative calculations}\label{AppA}
In performing the path integral over  the $\delta r$ and $y$ variables, we exploit the standard result for Gaussian functional integrals: 
\be
\begin{split}
\int\mathcal{D} \phi ~\, \textmd{exp} \left[ - \int_0^t d t' dt'' ~ \phi_i (t')  A_{ij}(t'-t'') \phi_j(t'')  +  \int_0^t d t' \,  B_i(t')  \phi_i(t')  + C  \right]&=\\
  ~\sqrt{ \frac{\pi}{\textmd{det} A_{ij} } } ~ \textmd{exp} \left[ ~\frac{1}{4} \, \int_0^t dt'dt'' \, \, B_i(t') \, A^{-1}_{ij}(t'-t'') \,  B_j(t'') + C ~ \right] &
\end{split}
\ee

The vibronic two-point functions $\Delta_{ij}(t'-t'')$ and $\mathcal{V}_{ij}(t-t')$, which enter in Eq.~(\ref{PIstep5}) are contracted from  the Green's functions of the $\hat L^\dag  \hat L$, $\hat L$ and $\hat L^\dag$ operators:
\be
\Delta_{ij}(t) ~ &\equiv&~ \LL[ \hat L^\dag  \hat L \RR]^{-1}(t) ~ = ~ \LL[ \, \LL( M \partial_{t}^2 \delta_{ij} + \gamma M \partial_{t} \delta_{ij} + \mathcal{H}_{ij} \RR) \, \LL( M \partial_{t}^2 \delta_{ij} - \gamma M \partial_{t} \delta_{ij} + \mathcal{H}_{ij} \RR) \, \RR]^{-1}, \\
\mathcal{V}_{ij}(t) ~ &\equiv& ~ \LL[ \hat L^\dag \RR]^{-1} (t) + \LL[ \hat L \RR]^{-1}(t) ~ = ~ 
\LL[ M \left( \partial_{t}^2 + \gamma \partial_{t} \right)\delta_{ij} + \mathcal{H}_{ij} \RR]^{-1} + 
\LL[ M \left( \partial_{t}^2 - \gamma \partial_{t} \right)\delta_{ij} + \mathcal{H}_{ij} \RR]^{-1} \, .              
\ee

In order to compute them it is convenient to consider the Fourier transform to frequency space. We also transform into the normal mode basis, by applying the unitary transformation $\hat U$ which diagonalizes the Hessian operators. We obtain: 
\bea 
 \widetilde{\Delta}_{ij}(\omega) 	& = & \frac{1}{M^2}   
	U^{\dag}_{in} \LL[ ~ \LL( \omega^2 - i \gamma \omega - \Omega_n \RR) \, \LL( \omega^2 + i \gamma \omega - \Omega_n \RR) ~ \RR]^{-1} U_{nj} \, , \\
 \widetilde{\mathcal{V}}_{ij}(\omega) 	& = & \frac{-1}{M}  
	U^{\dag}_{in} \LL[ ~ \LL( \omega^2 - i \gamma \omega - \Omega_n \RR)^{-1} + \LL( \omega^2 + i \gamma \omega - \Omega_n \RR)^{-1} ~ \RR] U_{nj} \, ,
\eea
where $\mathcal{H}_{ij}$ is the Hessian, and $\Omega_n$ are the corresponding normal modes. 
The expressions  (\ref{Delta}) and (\ref{DeltaV}) for $\Delta_{i j}(t)$ and $\mathcal{V}_{i j}(t)$ are obtained by Fourier transforming back to the time representation, taking the continuum limit for the Fourier sum. 

The following traces enter the derivation of the perturbative estimate (\ref{P1corr}):
\bea
\text{tr} \LL[\gamma^- \gamma^5 \gamma^+ \gamma^5 \RR] & = & N_{spin}  \\ 
- \text{tr} \LL[\gamma^- \gamma^0 \RR] = \text{tr} \LL[\gamma^+ \gamma^0 \RR] & = & N_{spin}  \\
-\text{tr} \LL[\gamma^- \gamma^0 \gamma^- \gamma^5  \gamma^+ \gamma^5 \RR] = \text{tr} \LL[\gamma^- \gamma^5 \gamma^+ \gamma^0 \gamma^+ \gamma^5 \RR]  
& = & N_{spin} \\
\text{tr} \LL[\gamma^- \gamma^5 \gamma^+ \gamma^0 \gamma^+ \gamma^0 \gamma^+ \gamma^5 \RR] = \text{tr} \LL[\gamma^+ \gamma^5 \gamma^- \gamma^0 \gamma^- \gamma^0 \gamma^- \gamma^5 \RR] & = & N_{spin}   
\eea
where $N_{spin}$ is the degeneracy number associated to the spin of the excitation and is equal to 2 for spin-1/2 fermions and 1 for spin-0 bosons. We recall that also the dimensionality and definition of the $\gamma$-matrixes depend on the spin of the propagating particle. In particular, for spin-0 bosons they reduce to  $ \gamma_0 = \tau_3$, and  $\gamma_5 = \tau_1$, where $\tau_3$ and $\tau_1$ are Pauli matrixes,

\section{Measuring Quantum Decoherence}
\label{quantumdecoherence}

In this appendix we review the proof that the ratio 
\be
R(t) = \text{Tr}[\hat\rho^2(t)]/\text{Tr}[\hat\rho(t)] \, ,
\ee
provides a measurement of the degree of decoherence of the system. 

We first consider  a pure state, denoted by the vector $|\chi\rangle$, and we represent the corresponding 
density operator with $\hat \rho = |\chi\rangle \langle \chi|$, so that $\text{Tr}[\hat\rho] = \langle \chi | \chi \rangle$. 

The operator $\hat \rho^2$ reads  $\hat \rho^2= |\chi\rangle \langle \chi|\chi\rangle \langle \chi|$, while its trace is 
\be
\text{Tr}[\hat \rho^2] = \langle \chi| \chi \rangle = (\text{Tr}[\hat\rho])^2,
\ee 
hence $R(t) = 1$. For a mixed state, there is no single state vector describing the system, and $R(t)< 1$.


\end{document}